# Accuracy, asymptotes, and applications of the Born and some other approximations to the calculation of the Mott scattering cross section, the Mott correction to the Bethe-Bloch formula, the primary atomic displacement cross section, and the energy-loss straggling

***Arkhutsik A.[1], Kats P.,[2] Voskresenskaya O.[3]***

[1]Brest State A.S. Pushkin University
[2]Brest State A.S. Pushkin University
[3]Joint Institute for Nuclear Research
e-mail: artboyarh@gmail.com
katspyotr@yandex.ru
voskr@jinr.ru

***Abstract.*** *The first, second and third-Born approximations of the Mott scattering cross section are considered. The relative error of all three Born approximations averaged by angles and energies is calculated for the first 30 elements of Mendeleev's periodic table of elements and also of the second and third-Born approximation for the first hundred elements of Mendeleev's table. The accuracy of the second and third-Born approximations and the LQZ method for calculating the normalized Mott scattering cross section on the nuclei of elements with Z = 1–6 are compared. The accuracy of the Born approximations for calculating the Mott correction in the Bethe–Bloch formula for Z = 1–27 for the second-Born approximation and Z = 1 – 100 for the third approximation is analyzed. An expression is obtained for the cross section of the primary displacement of the atom in the third-Born approximation. For iron, silver and lead, the cross section of the primary displacement of the atom for a number of electron energies is calculated. For a few examples, it was calculated that starting from what electron energy the difference in the cross section obtained using the asymptotic formula will be less than 1% compared to the value obtained using the McKinley–Feshbach formula. Accuracy of the Born approximations for calculating energy-loss straggling is analyzed.*

## Introduction

In 1911, Rutherford published an article [1] in which, within the framework of classical mechanics, he derived the famous Rutherford formula for the differential cross section for the scattering of charged particles by a point Coulomb center. For the scattering of electrons and positrons, the Rutherford cross section is expressed by the formula:

$$\sigma_R \equiv \left(\frac{d\sigma}{d\Omega}\right)_R = \left(\frac{Ze^2}{2mv^2}\right)^2 \frac{1}{\sin^4\left(\theta/2\right)}. \quad (1)$$

In 1928, the corresponding problem was solved in nonrelativistic quantum mechanics [2; 3]. By a lucky chance, classical mechanics for the Coulomb potential leads to the same formula as nonrelativistic quantum mechanics, which allowed Rutherford to create a nuclear model of the atom.

A little later, Mott found a solution to the problem within the framework of relativistic quantum mechanics [4; 5]

$$\sigma_M \equiv \left(\frac{d\sigma}{d\Omega}\right)_M = \left(\frac{\hbar}{mv}\right)^2 (1-\beta^2)\left(\frac{\xi^2 \mid F_M \mid^2}{\sin^2\left(\theta/2\right)} + \frac{\mid G_M \mid^2}{\cos^2\left(\theta/2\right)}\right), \quad (2)$$

where

$$F_M(\theta) = \frac{1}{2} i \sum_{l=0}^{\infty} (-1)^k [kC_M^{(k)} + (k+1)C_M^{(k+1)}] P_k(\cos\theta) = \sum_{l=0}^{\infty} F_M^{(k)} P_k(\cos\theta),$$

$$G_M(\theta) = \frac{1}{2} i \sum_{l=0}^{\infty} (-1)^k [k^2 C_M^{(k)} - (k+1)^2 C_M^{(k+1)}] P_k(\cos\theta) = \sum_{l=0}^{\infty} G_M^{(k)} P_k(\cos\theta),$$

$$C_M^{(k)} = -e^{-i\pi\rho_k} \frac{\Gamma(\rho_k - i\eta)}{\Gamma(\rho_k + 1 + i\eta)}, \ \eta = \frac{Z\alpha}{\beta}, \ \xi = \eta\sqrt{1-\beta^2}, \ \rho_k = \sqrt{k^2 - (Z\alpha)^2}, \ \alpha = \frac{e^2}{\hbar c},$$

$$F_M(\theta) = F_0(\theta) + F_1(\theta), \ G_M(\theta) = G_0(\theta) + G_1(\theta)$$

$$F_0(\theta) = \frac{1}{2} i \sum_{l=0}^{\infty} (-1)^k [kC_Z^{(k)} + (k+1)C_Z^{(k+1)}] P_k(\cos\theta),$$

$$G_0(\theta) = \frac{1}{2} i \sum_{l=0}^{\infty} (-1)^k [k^2 C_Z^{(k)} - (k+1)^2 C_Z^{(k+1)}] P_k(\cos\theta),$$

$$F_1(\theta) = \frac{1}{2} i \sum_{l=0}^{\infty} (-1)^k [kD^{(k)} + (k+1)D^{(k+1)}] P_k(\cos\theta),$$

$$G_1(\theta) = \frac{1}{2} i \sum_{l=0}^{\infty} (-1)^k [k^2 D^{(k)} - (k+1)^2 D^{(k+1)}] P_k(\cos\theta),$$

$$C_Z^{(k)} = -e^{-i\pi k} \frac{\Gamma(k - i\eta)}{\Gamma(k + 1 + i\eta)}, \quad D^{(k)} = C_M^{(k)} - C_Z^{(k)}.$$

Here

$$F_M(\theta)=\frac{1}{2}i\sum_{l=0}^{\infty}(-1)^k[kC_M^{(k)}\ +(k+1)C_M^{(k+1)}]P_k(\cos\theta)=\sum_{l=0}^{\infty}F_M^{(k)}P_k(\cos\theta)\,,$$

$$G_M(\theta)=\frac{1}{2}i\sum_{l=0}^{\infty}(-1)^k[k^2C_M^{(k)}-(k+1)^2C_M^{(k+1)}]P_k(\cos\theta)=\sum_{l=0}^{\infty}G_M^{(k)}P_k(\cos\theta),$$

$$C_M^{(k)}=-e^{-i\pi\rho_k}\frac{\Gamma(\rho_k-i\eta)}{\Gamma(\rho_k+1+i\eta)},\ \eta=\frac{Z\alpha}{\beta},\ \xi=\eta\sqrt{1-\beta^2},\ \rho_k=\sqrt{k^2-(Z\alpha)^2},\ \alpha=\frac{e^2}{\hbar c},$$

$$F_M(\theta)=F_0(\theta)+F_1(\theta)\,,\ G_M(\theta)=G_0(\theta)+G_1(\theta)$$

$$F_0(\theta)=\frac{1}{2}i\sum_{l=0}^{\infty}(-1)^k[kC_Z^{(k)}\ +(k+1)C_Z^{(k+1)}]P_k(\cos\theta),$$

$$G_0(\theta)=\frac{1}{2}i\sum_{l=0}^{\infty}(-1)^k[k^2C_Z^{(k)}-(k+1)^2C_Z^{(k+1)}]P_k(\cos\theta),$$

$$F_1(\theta)=\frac{1}{2}i\sum_{l=0}^{\infty}(-1)^k[kD^{(k)}\ +(k+1)D^{(k+1)}]P_k(\cos\theta),$$

$$G_1(\theta)=\frac{1}{2}i\sum_{l=0}^{\infty}(-1)^k[k^2D^{(k)}-(k+1)^2D^{(k+1)}]P_k(\cos\theta),$$

$$C_Z^{(k)}=-e^{-i\pi k}\frac{\Gamma(k-i\eta)}{\Gamma(k+1+i\eta)},\quad D^{(k)}=C_M^{(k)}-C_Z^{(k)}\,.$$

Here α is the fine structure constant, Γ is the gamma function, and $P_k$ are the Legendre polynomials. The functions $F_0(\theta)$ and $G_0(\theta)$ can be written as

$$F_0(\theta)=\frac{i}{2}\frac{\Gamma(1-i\eta)}{\Gamma(1+i\eta)}\sin^{2i\eta}\left(\frac{\theta}{2}\right),\ G_0(\theta)=-i\eta\frac{F_0(\theta)}{\tan^2(\theta/2)}\,. \tag{3}$$

Formula (2) is called “the exact formula for the differential scattering cross section”.

The first numerical summation of these series was performed by Mott [5]. Starting from this work, in similar calculations they began to introduce a quantity equal to the ratio to the modified Rutherford cross section ($\tilde{\sigma}_R$),

$$R(\theta)=\sigma_M/\tilde{\sigma}_R,\quad \tilde{\sigma}_R=\sigma_R(1-\beta^2),\ \beta=\frac{v}{c}, \tag{4}$$

or the normalized Mott cross section (NMS). It can be represented as:

$$R_M(\theta)=\frac{4\sin^2(\theta/2)}{\eta^2}\left[\xi^2\,|\,F\,|^2+\tan^2\left(\frac{\theta}{2}\right)|\,G\,|^2\right]. \tag{5}$$

Since the exact Mott scattering cross section (2) and the NMS (5) involve slowly converging series, their application is complex. Therefore, the use of analytical approximations becomes important.

One way to obtain such approximations is to expand the exact NMS into a power series in αZ. These are called Born approximations of the Mott scattering cross section [6]. The first-Born approximation was obtained by Mott himself [5]:

$$R_B(\theta) = 1 - \beta^2 \sin^2\left(\frac{\theta}{2}\right). \tag{6}$$

The second Born approximation was obtained by McKinley and Feshbach [7]:

$$R_{MF}(\theta) = R_B + \pi\alpha\beta Z \sin\left(\frac{\theta}{2}\right)\left[1 - \sin\left(\frac{\theta}{2}\right)\right]. \tag{7}$$

The third-Born approximation was obtained by Johnson, Weber, and Mullin [8]:

$$\begin{aligned} R_{JWM} = R_{MF} + (\alpha Z)^2 \sin\left(\frac{\theta}{2}\right)\Bigg\{ & L_2\left[1-\sin^2\left(\frac{\theta}{2}\right)\right] - 4L_2\left[1-\sin\left(\frac{\theta}{2}\right)\right] + 2\sin\left(\frac{\theta}{2}\right)\ln^2\left[\sin\left(\frac{\theta}{2}\right)\right] \\ & + \frac{\pi^2}{2}\left[1-\sin\left(\frac{\theta}{2}\right)\right] + \frac{\pi^2}{6}\sin\left(\frac{\theta}{2}\right) + \beta^2\sin\left(\frac{\theta}{2}\right)\left(L_2\left[1-\sin^2\left(\frac{\theta}{2}\right)\right] + \frac{\sin^2(\theta/2)\ln^2[\sin(\theta/2)]}{1-\sin^2(\theta/2)} + \right. \\ & \left. + \frac{\pi^2}{4}\frac{1-\sin(\theta/2)}{1+\sin(\theta/2)} - \frac{\pi^2}{6}\right)\Bigg\}, \end{aligned} \tag{8}$$

where $L_2$ denotes the Euler dilogarithm:

$$L_2(x) = -\int_0^x \frac{\ln(1-y)}{y}dy.$$

The first-Born approximation is widely used in particle physics [9]. The Bethe–Bloch formula for the energy loss of charged particles in matter was derived in this approximation [10].

The second-Born approximation is often used in the analysis of radiation damage to matter by electrons [11–15]. This approximation yields a simple formula for the cross section for the primary displacement of an atom by an electron [16]. The Mott correction to the Bethe–Bloch formula was also derived in the second Born approximation [17].

The third-Born approximation is used less frequently due to its rather complex expression. Morgan and Ebi [18] derived the Mott correction to the Bethe–Bloch formula in the third Born approximation.

The accuracy of the Born approximations has been estimated for various special cases by the authors of the papers [7, 8] and by researchers of [18–22]. One of the most extensive comparisons of calculations in the second-Born approximation and exact calculations is given by Oen [23]. McKinley and Feshbach already estimated the applicability limits of their method as $Z\alpha \leq 0.2$, i.e., $Z \leq 27$. However, there are a number of studies in which the second Born approximation is applied to cases of $Z > 30$ [24–28]. The monograph [29, p. 15] claims that the calculation using (7) for $Z \leq 40$ does not exceed 1%.

In this paper, we analyze the accuracy of the Born approximations for calculating the Mott scattering cross section and the Mott correction for a wide Z and β range. We calculate the atomic displacement cross section for some examples and analyze the applicability conditions for the high-energy approximations of the displacement cross section. We also analyze the accuracy of the Born approximation in calculating the energy-loss straggling.

## Verification of the Accuracy of the Born Approximations for the Mott Scattering Cross Section

To characterize the accuracy of approximate methods, we will use the concept of the mean relative error ER, borrowed from [30]:

$$ER = \sqrt{\frac{\sum_{i=1}^{36}[R(\theta_i) - R^{EXACT}(\theta_i)]^2}{\sum_{i=1}^{36}[R^{EXACT}(\theta_i)]^2}}. \quad (9)$$

The arithmetic mean of the energy relative error was calculated for 15 energy values, ranging from 5 keV to 10 MeV. The results are presented in Table 1 and Figure 1.

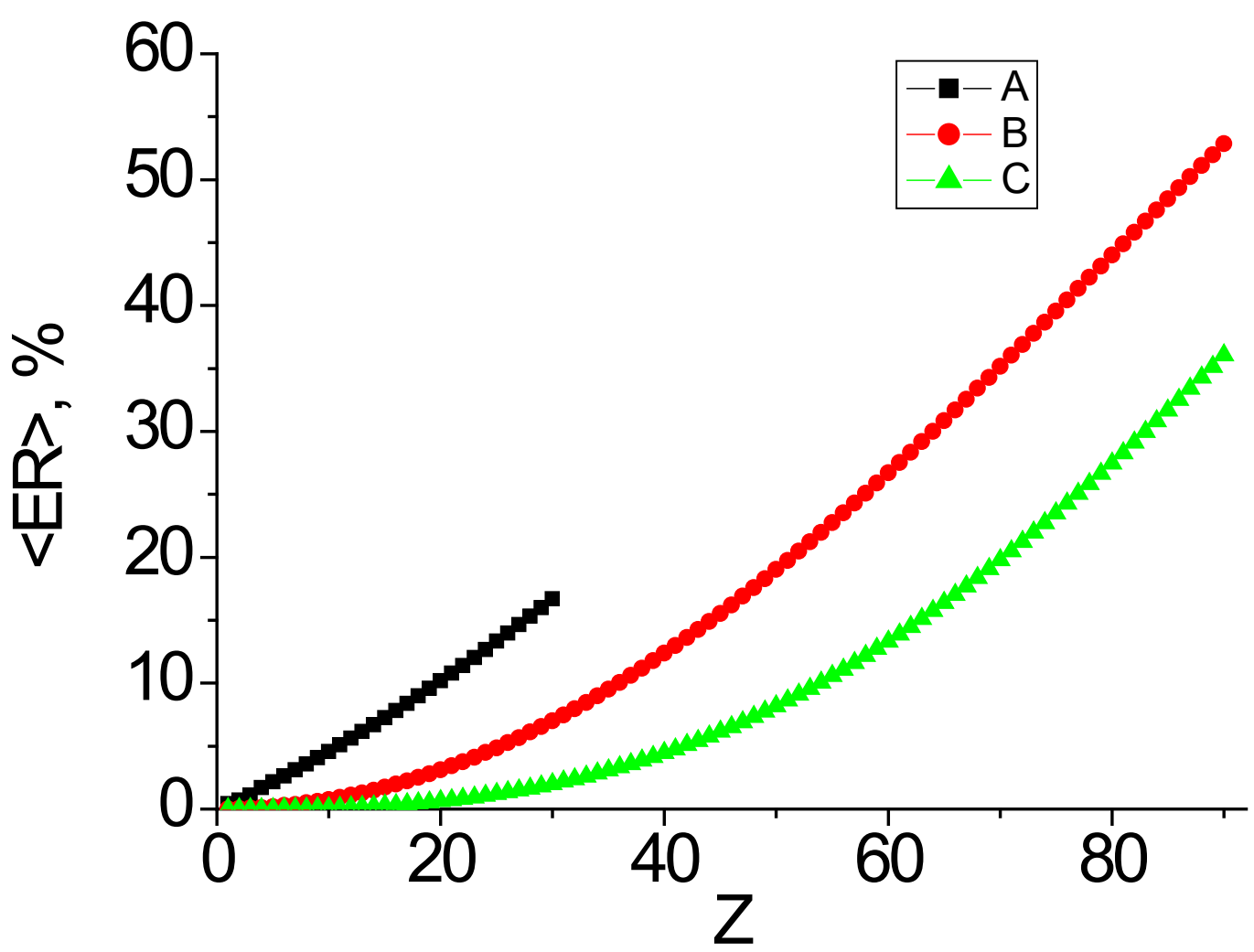

**Figure 1.** The arithmetic mean of the relative errors: A is $\langle ER \rangle_B$, B is $\langle ER \rangle_{MF}$, and C is $\langle ER \rangle_{JWM}$.

**Table 1. The arithmetic mean of the relative error $\langle ER \rangle$ for 15 energy values, from 5 keV to 10 MeV**

| Z | 1 | 2 | 3 | 4 | 5 | 6 | 7 | 8 | 9 |
|---|---|---|---|---|---|---|---|---|---|
| $\langle ER \rangle_B$, % | 0,41 | 0,83 | 1,3 | 1,7 | 2,2 | 2,6 | 3,1 | 3,6 | 4,1 |
| $\langle ER \rangle_{MF}$, % | $7{,}2\cdot10^{-3}$ | 0,029 | 0,066 | 0,12 | 0,18 | 0,27 | 0,37 | 0,48 | 0,61 |
| $\langle ER \rangle_{JWM}$, % | $9{,}0\cdot10^{-5}$ | $7{,}1\cdot10^{-4}$ | $2{,}4\cdot10^{-3}$ | $5{,}6\cdot10^{-3}$ | 0,011 | 0,019 | 0,030 | 0,044 | 0,062 |
| Z | 10 | 11 | 12 | 13 | 14 | 15 | 16 | 17 | 18 |
| $\langle ER \rangle_B$, % | 4,6 | 5,1 | 5,6 | 6,2 | 6,7 | 7,3 | 7,8 | 8,4 | 9,0 |
| $\langle ER \rangle_{MF}$, % | 0,75 | 0,91 | 1,1 | 1,3 | 1,5 | 1,7 | 2,0 | 2,2 | 2,5 |
| $\langle ER \rangle_{JWM}$, % | 0,085 | 0,11 | 0,14 | 0,18 | 0,225 | 0,27 | 0,33 | 0,39 | 0,46 |
| Z | 19 | 20 | 21 | 22 | 23 | 24 | 25 | 26 | 27 |
| $\langle ER \rangle_B$, % | 9,6 | 10,2 | 10,8 | 11,4 | 12,0 | 12,7 | 13,3 | 14,0 | 14,6 |
| $\langle ER \rangle_{MF}$, % | 2,8 | 3,1 | 3,4 | 3,8 | 4,1 | 4,5 | 4,9 | 5,3 | 5,7 |
| $\langle ER \rangle_{JWM}$, % | 0,54 | 0,63 | 0,72 | 0,83 | 0,94 | 1,1 | 1,2 | 1,3 | 1,5 |
| Z | 28 | 29 | 30 | 31 | 32 | 33 | 34 | 35 | 36 |
| $\langle ER \rangle_B$, % | 15,3 | 16,0 | 16,7 | | | | | | |
| $\langle ER \rangle_{MF}$, % | 6,1 | 6,5 | 7,0 | 7,5 | 8,0 | 8,5 | 9,0 | 9,5 | 10,1 |
| $\langle ER \rangle_{JWM}$, % | 1,6 | 1,8 | 2,0 | 2,2 | 2,4 | 2,6 | 2,8 | 3,1 | 3,3 |
| Z | 37 | 38 | 39 | 40 | 41 | 42 | 43 | 44 | 45 |
| $\langle ER \rangle_{MF}$, % | 10,6 | 11,2 | 11,8 | 12,4 | 13,0 | 13,6 | 14,2 | 14,9 | 15,5 |
| $\langle ER \rangle_{JWM}$, % | 3,6 | 3,9 | 4,1 | 4,4 | 4,8 | 5,1 | 5,4 | 5,8 | 6,1 |
| Z | 46 | 47 | 48 | 49 | 50 | 51 | 52 | 53 | 54 |
| $\langle ER \rangle_{MF}$, % | 16,2 | 16,9 | 17,6 | 18,3 | 19,0 | 19,7 | 20,5 | 21,2 | 22,0 |
| $\langle ER \rangle_{JWM}$, % | 6,5 | 6,9 | 7,3 | 7,7 | 8,2 | 8,6 | 9,1 | 9,6 | 10,1 |
| Z | 55 | 56 | 57 | 58 | 59 | 60 | 61 | 62 | 63 |
| $\langle ER \rangle_{MF}$, % | 22,7 | 23,5 | 24,3 | 25,1 | 25,9 | 26,7 | 27,5 | 28,3 | 29,2 |
| $\langle ER \rangle_{JWM}$, % | 10,6 | 11,1 | 11,6 | 12,2 | 12,7 | 13,3 | 13,9 | 14,5 | 15,1 |
| Z | 64 | 65 | 66 | 67 | 68 | 69 | 70 | 71 | 72 |
| $\langle ER \rangle_{MF}$, % | 30,0 | 30,9 | 31,7 | 32,6 | 33,4 | 34,3 | 35,2 | 36,0 | 36,9 |
| $\langle ER \rangle_{JWM}$, % | 15,7 | 16,4 | 17,0 | 17,7 | 18,4 | 19,1 | 19,8 | 20,5 | 21,2 |
| Z | 73 | 74 | 75 | 76 | 77 | 78 | 79 | 80 | 81 |
| $\langle ER \rangle_{MF}$, % | 37,8 | 38,7 | 39,6 | 40,4 | 41,3 | 42,2 | 43,1 | 44,0 | 44,9 |
| $\langle ER \rangle_{JWM}$, % | 22,0 | 22,7 | 23,5 | 24,3 | 25,0 | 25,8 | 26,6 | 27,5 | 28,3 |
| Z | 82 | 83 | 84 | 85 | 86 | 87 | 88 | 89 | 90 |
| $\langle ER \rangle_{MF}$, % | 45,8 | 46,7 | 47,6 | 48,5 | 49,3 | 50,2 | 51,1 | 52,0 | 52,9 |
| $\langle ER \rangle_{JWM}$, % | 29,1 | 30,0 | 30,8 | 31,7 | 32,5 | 33,4 | 34,3 | 35,1 | 36,0 |

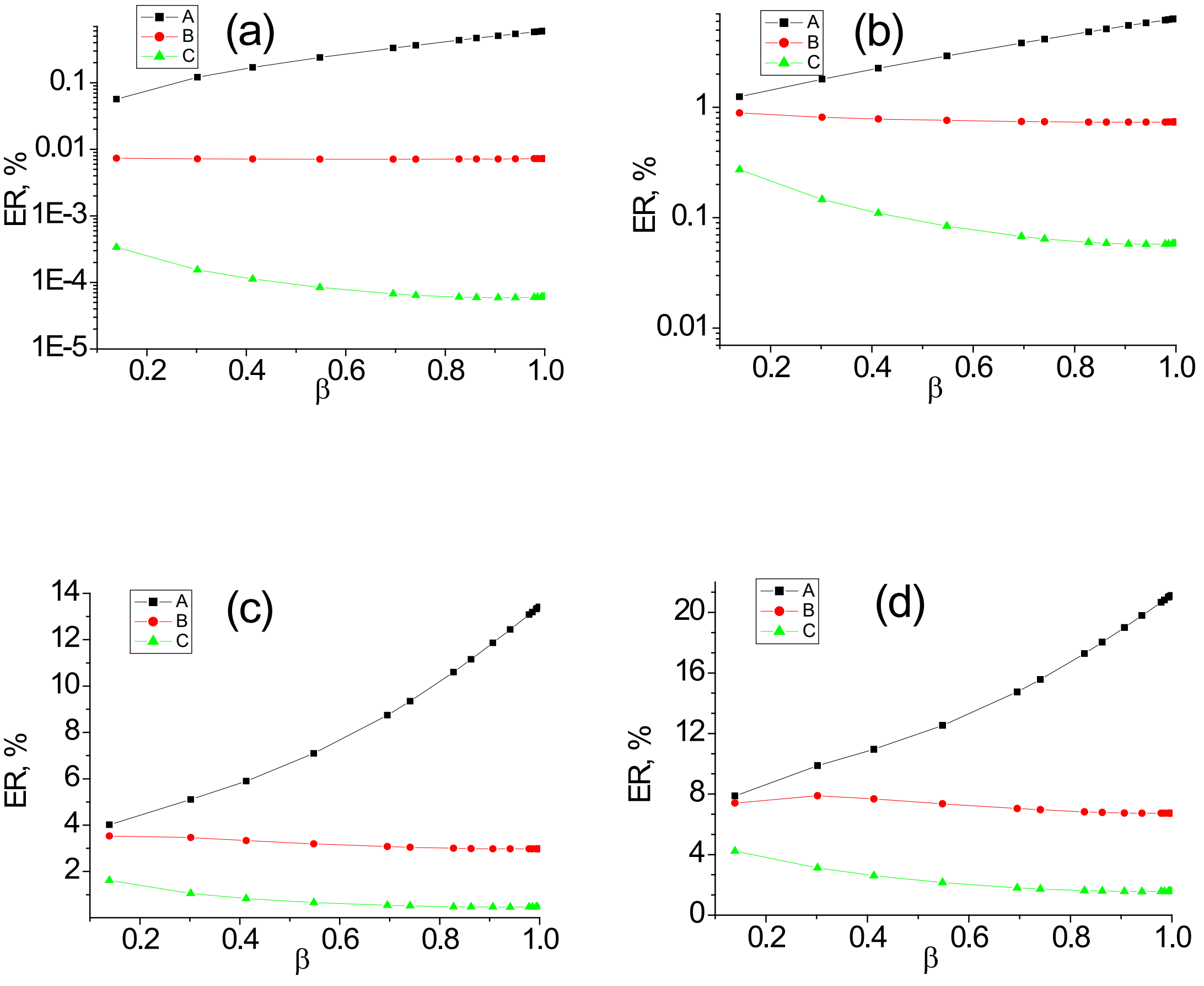

**Figure 2.** ER as a function of relative velocity: A is $\langle ER\rangle_B$, B is $\langle ER\rangle_{MF}$, and C is $\langle ER\rangle_{JWM}$ for elements with Z = 1 (a), 10 (b), 20 (c), and 30 (d).

The Figure 2 shows that the error in the first Born approximation increases with increasing velocity, the error in the second-Born approximation depends weakly on velocity, and the error in the third-Born approximation decreases with increasing velocity. Accuracy increases with the approximation number.

$\langle ER\rangle_B$ exceeds 1% for lithium, 3% for nitrogen, 5% for sodium, 10% for calcium, and 15% for nickel. It exceeds 1% for magnesium, 3% for calcium, and 5% for iron. The McKinley–Feshbach approximation is generally considered applicable up to Z = 27. According to our calculations, it reaches 5.7%. It exceeds 10% for krypton, 15% for rhodium, and 20% for tellurium. $\langle ER\rangle_{JWM}$ exceeds 1% for chromium, and reaches 2% for zinc, exceeds 3% for bromine and 5% for molybdenum, and 10% for xenon.

The relative error in calculating the NMS, averaged over angles and energies, exceeds 10% for the first-Born approximation, starting with Z = 20, for the second-Born approximation, with Z = 36, and for the third-Born approximation, with Z = 54.

The Figure 3 shows the ER(β) dependence for elements with Z = 36, 42, 48, and 54. It is evident from this Figure that the error in the second-Born approximation initially increases with increasing velocity, reaches a maximum in the region of β = 0,3–0,4, and then it decreases.

According to our calculations, for Z = 40 ER exceeds 12%, at 25 keV ER = 13.7%. Moreover, the relative error in calculating the normalized Mott cross section for this energy exceeds 20% at scattering angles greater than 150°. This refutes the statement given in [29] that the error in the second-Born approximation for Z ≤ 40 does not exceed 1%. Interestingly, up to 87° the relative error is less than 10%, and up to 60° the error of the second-Born approximation is less than that of the third one. It should be noted that the normalized cross section itself in this range of angles and energies differs little from unity (by less than 10%).

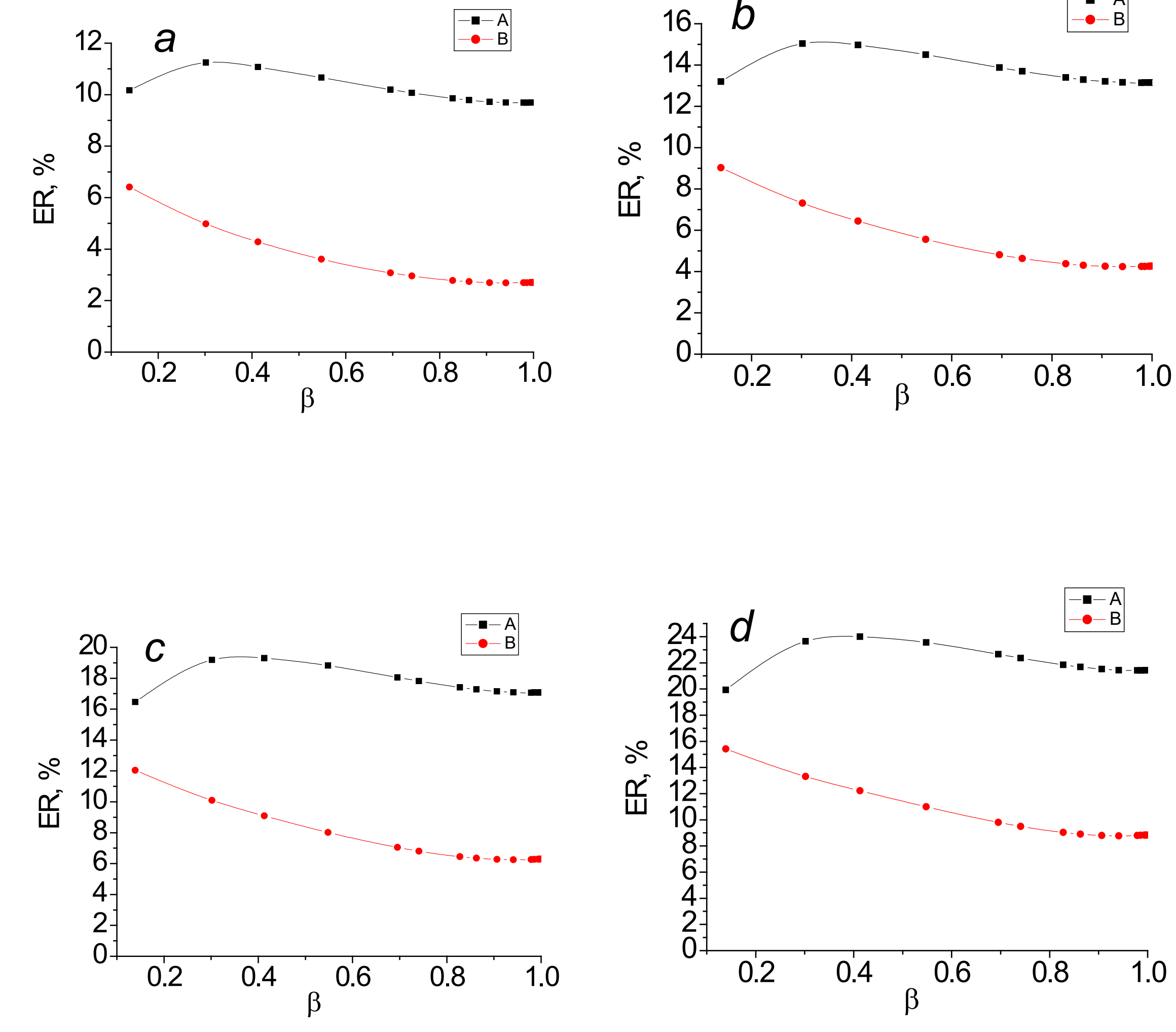

**Figure 3.** ER as a function of relative velocity: A anb B are $ER_{MF}$, $ER_{JWM}$, correspondingly, for Z = 36 (a), 42 (b), 48 (c), 54 (d)

In [8], according to the graph, a very noticeable difference is observed between the normalized scattering cross section calculated numerically and that obtained in the third-Born approximation for tin at an electron energy of 0,1 MeV.

Figure 4 shows the graph from [8] and our graph. The fragment of the graph from [8] also shows the graphs for positrons. As can be seen from our graph, the third-Born approximation does not actually produce this error. It can be assumed that [8] contains an error in the graph of the exact normalized Mott cross section. Figure 5 shows the dependence of ER on the relative velocity β = v/c for Z = 60, 70, 80, and 90. For larger Z, a maximum on this graph becomes noticeable. As Z increases, the position of the error maximum shifts toward higher velocities. For Z = 22, the maximum ER value is reached around β = 0,2, for Z = 60, around β = 0,42, and for Z = 90, around β = 0,55.

Figure 6 shows graphs of the dependence of the NMS R(θ) on for different energies for Z = 27, up to which the second Born approximation is considered applicable.

Up to 25 keV, the accuracy of the nonrelativistic approximation (Rutherford cross section) is higher than that of the first- and second-Born approximations.

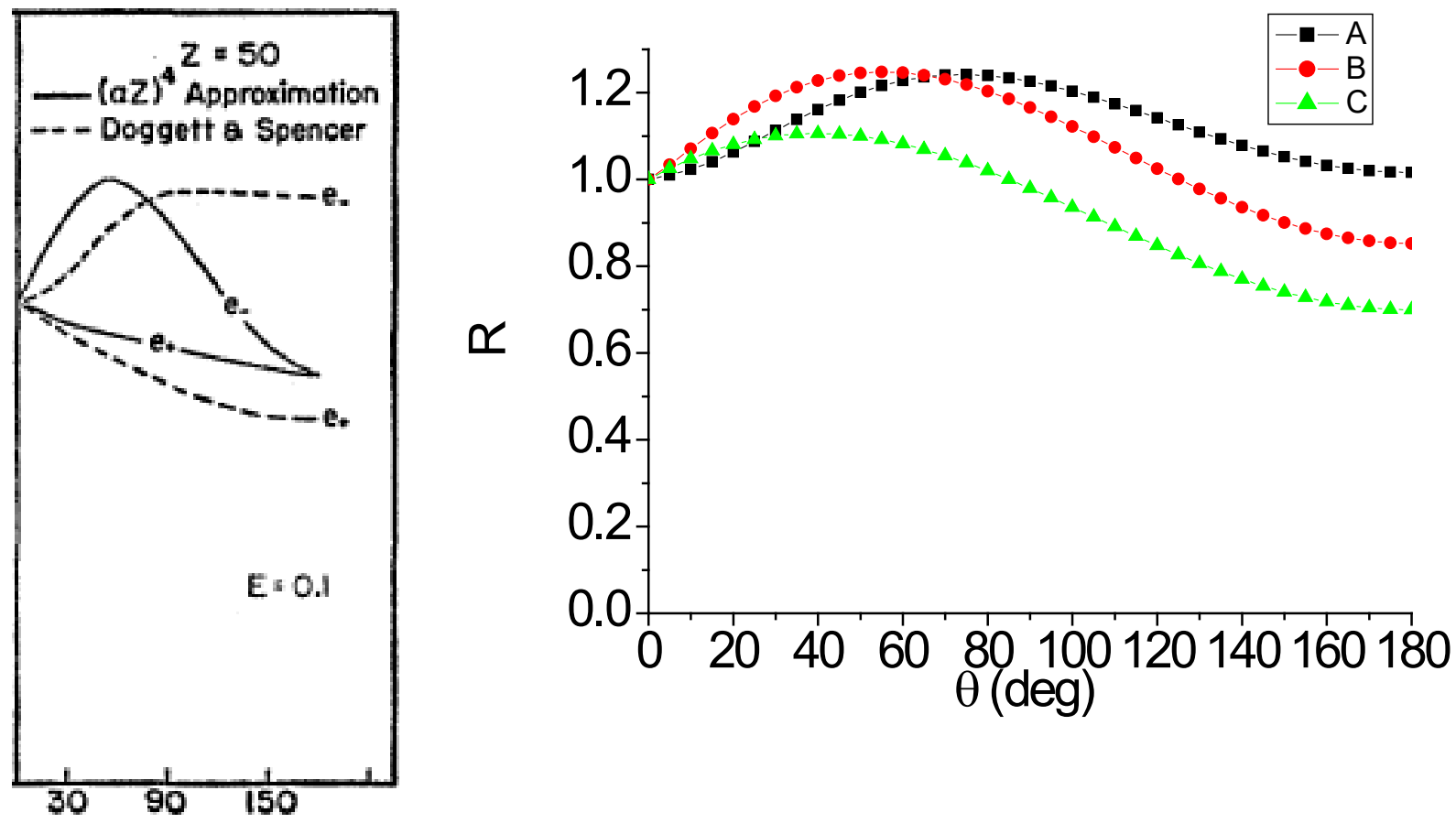

**Figure 4.** Comparison of the numerical calculation and the third Born approximation for tin at an electron energy of 0.1 MeV according to [8] and our calculations. A is the result of the numerical calculation of the NMS, B is the third-Born approximation, C is the second-Born approximation.

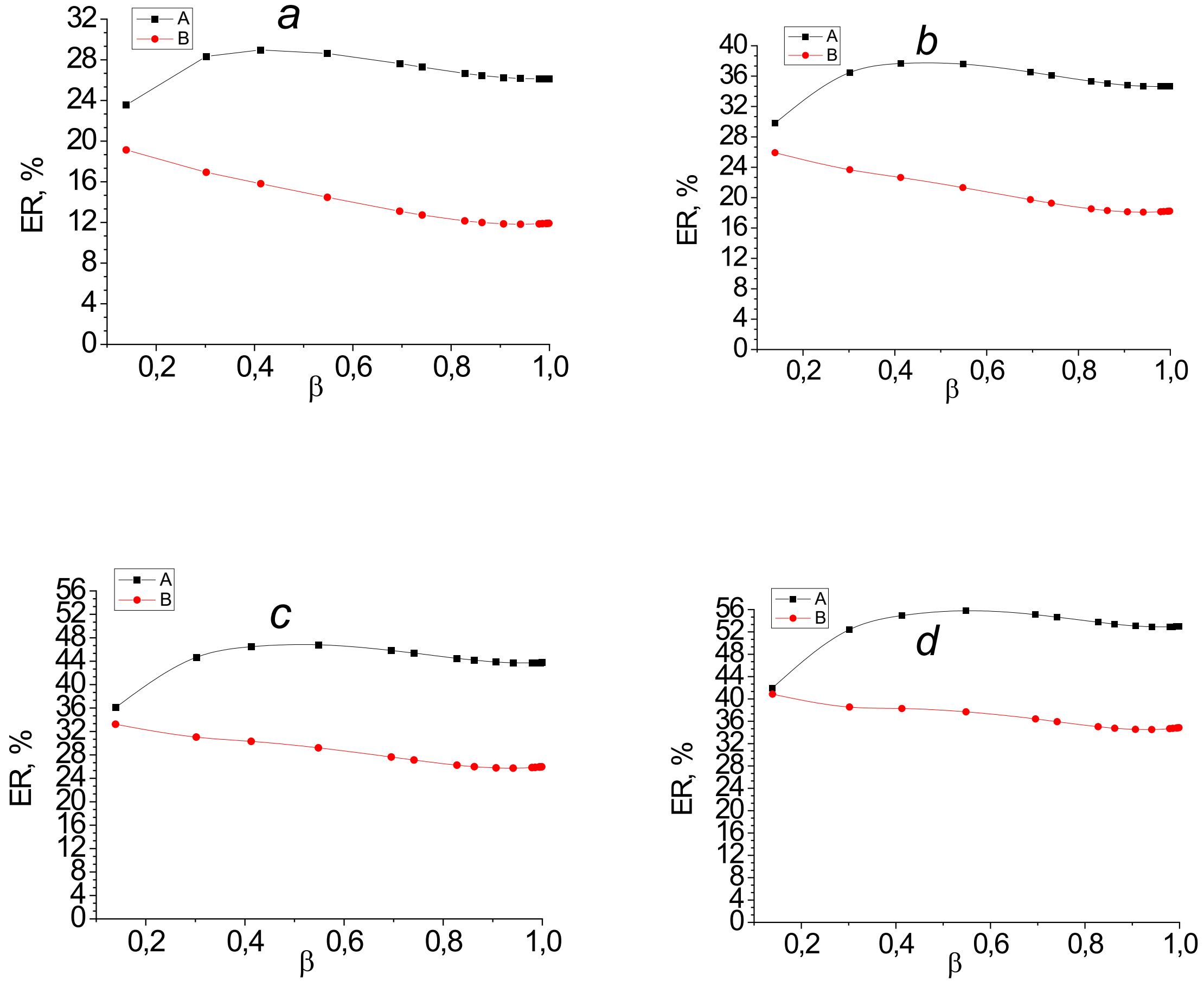

**Figure 5.** ER as a function of relative velocity: A anb B are $ER_{MF}$ and $ER_{JWM}$, respectively, for Z = 60 (a), 70 (b), 80 (c), and 90 (d).

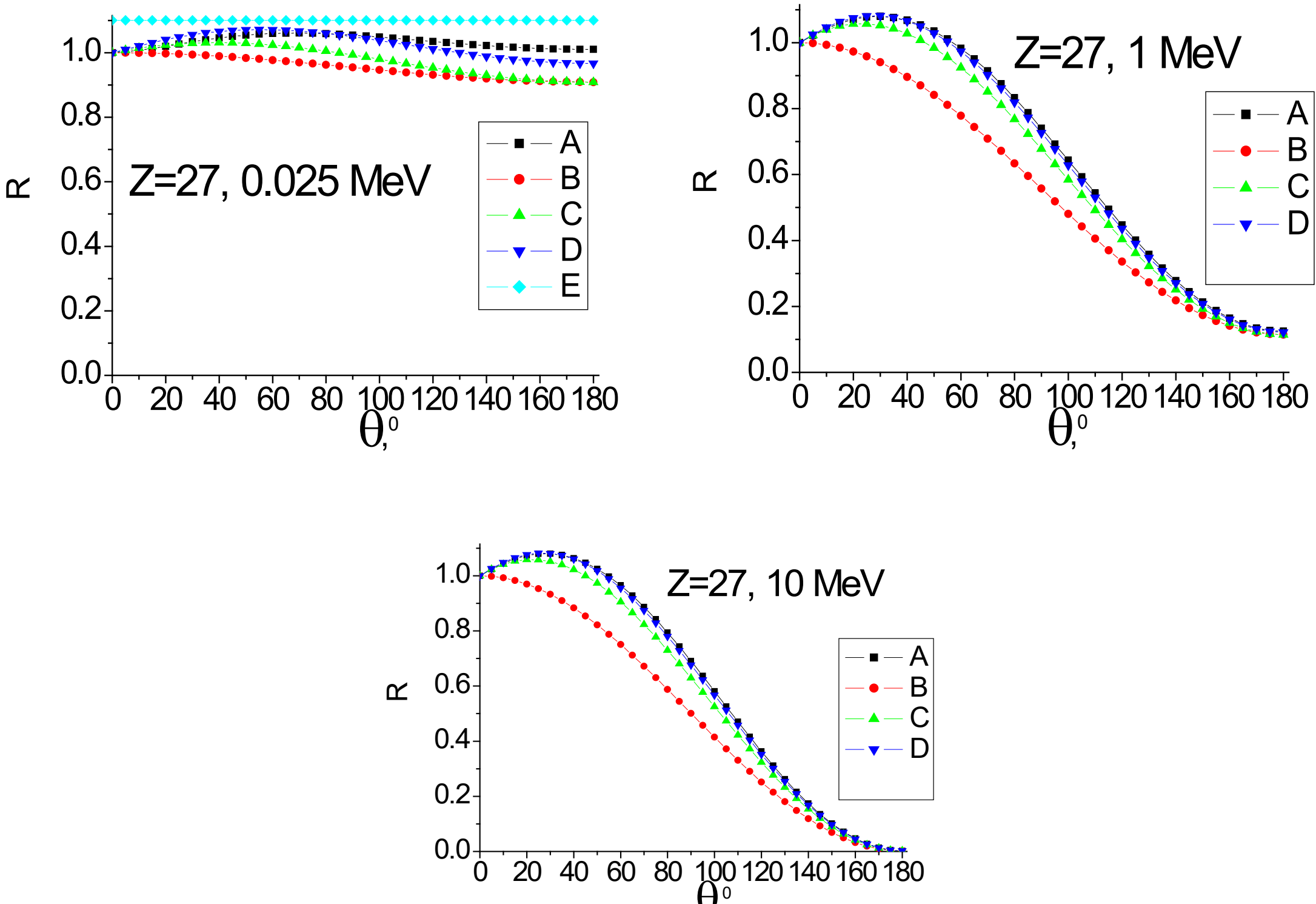

**Figure 6.** Nonrelativistic cross section R(θ) as a function of the scattering angle: A is numerical calculation, B is first-Born approximation, C is second-Born approximation, D is third-Born approximation, E is nonrelativistic approximation for the scattering of 0,025 MeV, 1 MeV, and 10 MeV electrons by nuclei with Z = 27.

The results for arithmetic mean of the relative error over energies both for positrons and electrons are presented in Figure 7.

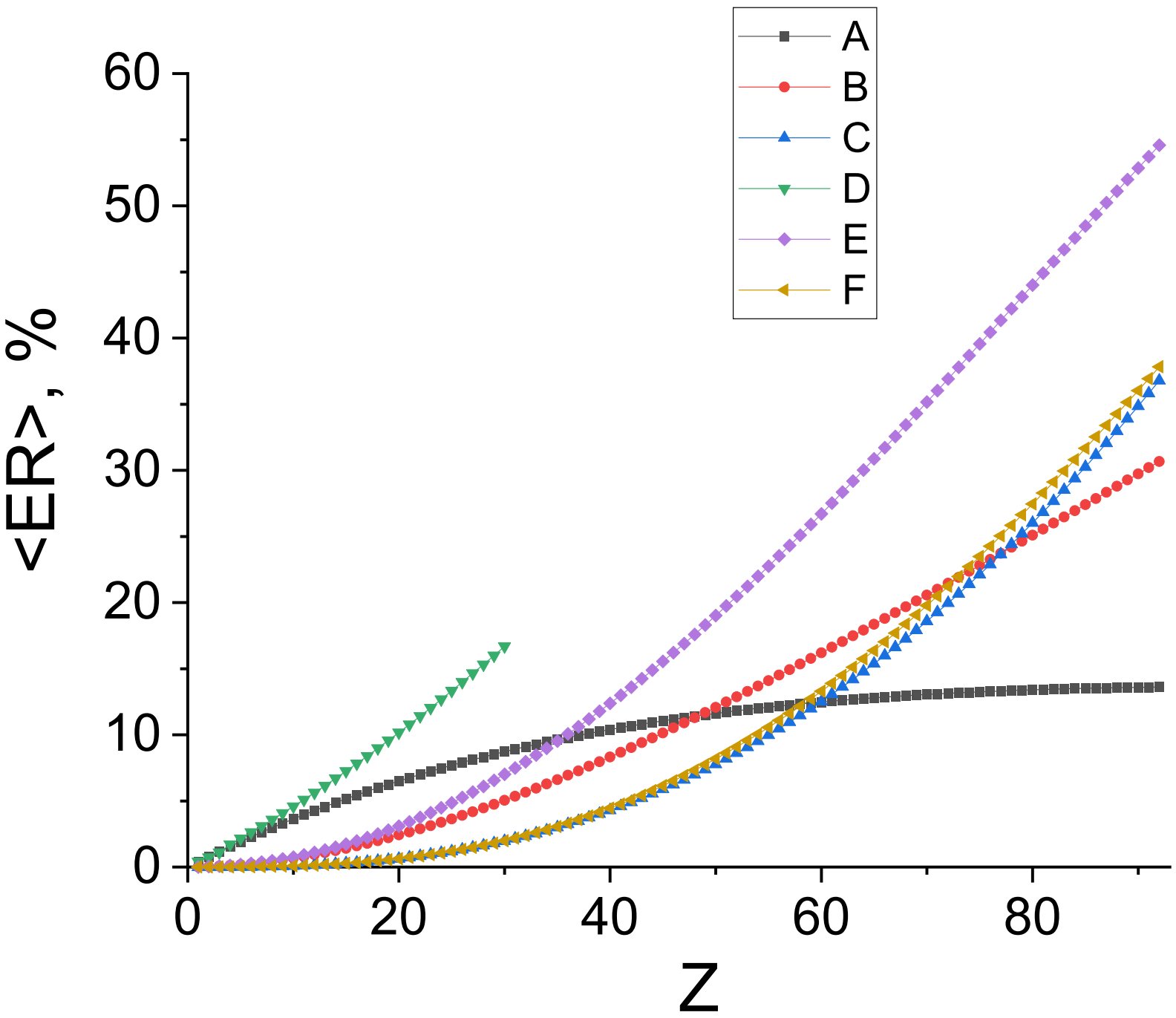

**Figure 7.** Arithmetic mean of the relative error *<ER>, %*. From A to C *<ER>* is calculuted for the positron scattering, where A is $\langle ER \rangle_B$, B is $\langle ER \rangle_{MF}$, and C is $\langle ER \rangle_{JWM}$. From D to F it is calculuted for electron scattering, where D is $\langle ER \rangle_B$, E is $\langle ER \rangle_{MF}$, and F is $\langle ER \rangle_{JWM}$.

The errors in the first-Born approximation for electrons and positrons are virtually identical up to Z = 6. Thereafter, with increasing Z, the error for electrons increases very rapidly, reaching almost 17% at Z = 30. For positrons, the error increases slowly, remaining less than 14% even at Z = 92.

The average error in the second-Born approximation for electrons is greater than for positrons. The difference increases with Z. For Z = 92, it is approximately 31% for positrons, and approximately 55% for electrons. However, it should be kept in mind that this is true only for the average error. For positrons, the second Born approximation can lead to meaningless results: zero or negative differential scattering cross sections. For example, for Z = 92 and a positron energy of 10 MeV, the differential scattering cross section in the second-Born approximation is less than zero for scattering angles from approximately 137 to 152 degrees.

It is quite interesting that, unlike electron scattering, for positrons, starting at Z = 49, the average error of the first-Born approximation becomes smaller than that of the second one and similarly, starting at Z = 60, it becomes smaller than that of the third. Consequently, the average error of the second-Born approximation becomes smaller than that of the third one, starting at Z = 78.

We present plots of the β dependence for characteristic values of Z.

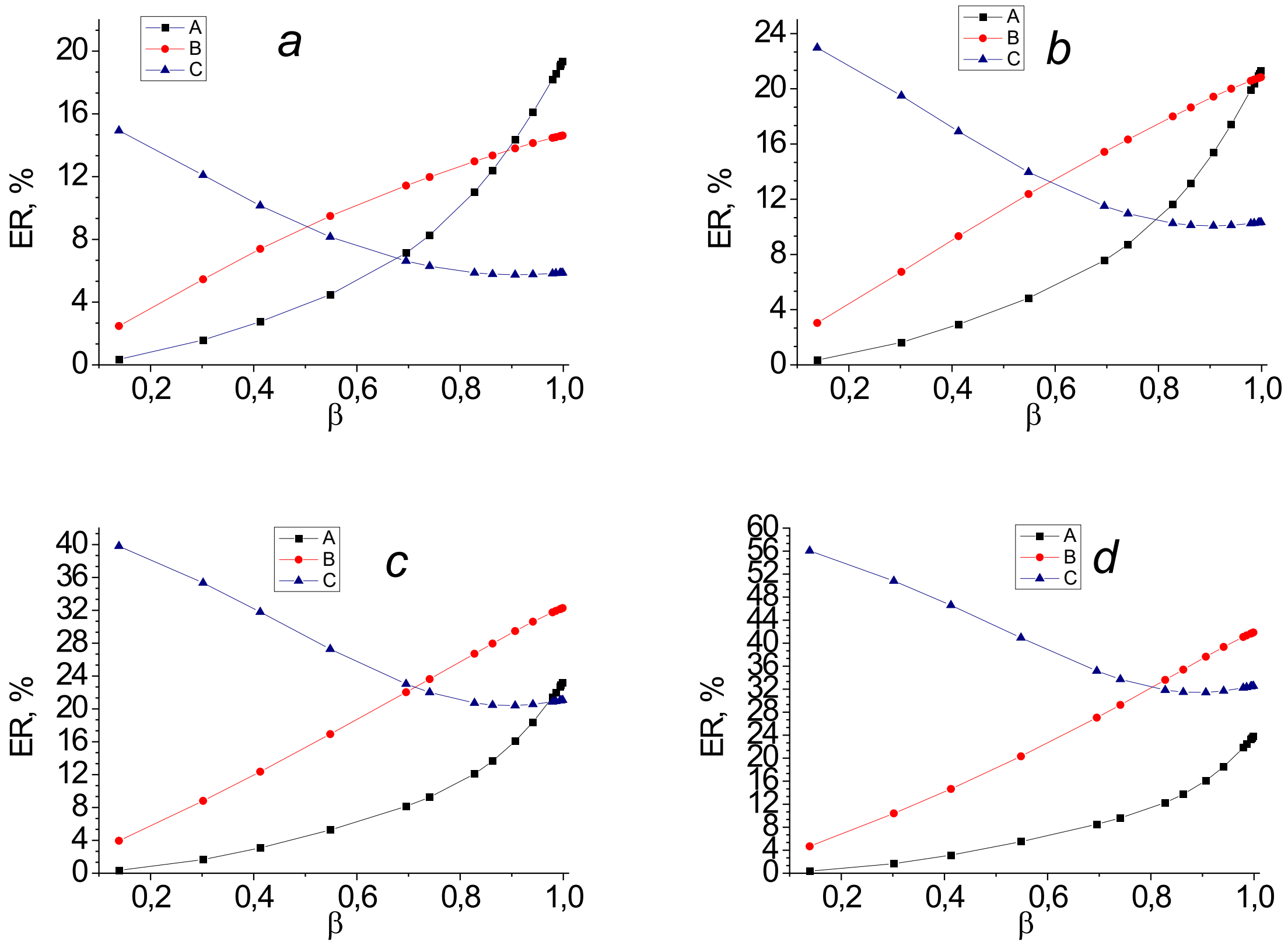

**Figure 8.** *ER* as a function of a relative velocity during positron scattering: for Z = 49 (*a*), 60 (*b*), 78 (*c*), 92 (*d*); A, B, and C are $\langle ER \rangle_{B}$, $\langle ER \rangle_{MF}$, and $\langle ER \rangle_{JWM}$, respectively.

As with electrons, the error in the first-Born approximation increases with velocity. The error in the second Born approximation also increases clearly with velocity. For electrons, the increase was weaker. The error in the third-Born approximation decreases with velocity and passes through a minimum at high velocities.

For Z = 49, at velocities below approximately 0.9 c, the first-Born approximation is more accurate than the second and up to 0.7 c, it is more accurate than the third. The second approximation is more accurate than the third up to approximately 0.5 c.

For Z = 60, the first-Born approximation becomes more accurate than the third starting at approximately 0.8 c and becomes worse than the second approximation at subluminal velocities – above 0.99 c. For Z = 78 and Z = 92, the first approximation is more accurate than the second at all speeds, and for Z = 92, it is more accurate than the third one.

**Comparison of the accuracy of the Born approximations and the LQZ method for calculating the normalized Mott scattering cross section on nuclei of light elements**

In 1995, a new approximation based on an analytical fit to the exact Mott cross section was proposed in [30]. We call this approximation the LQZ approximation. For each element, 30 coefficients are calculated using the least-squares method. As the authors of [30] showed, the LQZ approximation yields a small error. As the element's atomic number increases, the error of the Born approximations increases, and for large Z, the LQZ method proves preferable.

In this section, we consider the question of whether there are elements for which the accuracy of the Born approximations exceeds that of the LQZ method.

The average relative error ER was calculated for 26 velocities from 0.1 *c* to 0.999 *c* for the MF, JWM, and LQZ methods, as well as the averaging of this value over the velocities. The coefficients for the LQZ method were taken from [30] and [31]. The results are presented in Table 2.

**Table 2.** Arithmetic mean of the relative error of the relative error *<ER>, %.*

| Z | 1 | 2 | 3 | 4 | 5 | 6 |
|---|---|---|---|---|---|---|
| $\langle ER\rangle_{MF}$, % | $7{,}2\cdot10^{-3}$ | 0,029 | 0,066 | 0,12 | 0,19 | 0,27 |
| $\langle ER\rangle_{JWM}$, % | $1{,}1\cdot10^{-4}$ | $8{,}7\cdot10^{-4}$ | $2{,}9\cdot10^{-3}$ | $6{,}8\cdot10^{-3}$ | 0,013 | 0,023 |
| $\langle ER\rangle_{LQZ}$, % [30] | $1{,}2\cdot10^{-2}$ | $5{,}9\cdot10^{-3}$ | $4{,}0\cdot10^{-3}$ | $4{,}0\cdot10^{-3}$ | $3{,}4\cdot10^{-3}$ | $4{,}1\cdot10^{-3}$ |
| $\langle ER\rangle_{LQZ}$, % [31] | $8{,}0\cdot10^{-5}$ | $3{,}0\cdot10^{-4}$ | $7{,}4\cdot10^{-4}$ | $1{,}4\cdot10^{-3}$ | $2{,}4\cdot10^{-3}$ | $3{,}6\cdot10^{-3}$ |

For JWM, the error is lower than the error in LQZ with the coefficients from [30]. The error in LQZ with the coefficients from [31] is lower than the error in the Born approximations for all elements. For hydrogen, the error in LQZ using the coefficients from [30] is anomalously large and even exceeds the error in MF.

With very small errors, the accuracy of the calculation of the normalized Mott scattering cross section can influence the result. To calculate the exact value of the NMS, we used the summation of Mott series according to the method proposed in [22] up to L = 150. To check the influence of the number of terms in the sums, we calculated the error of the JWM method for Z = 1, β = 0,999 (at this speed, the minimum error was obtained) at L = 150 and L = 300. In the first case, $\langle ER\rangle_{JWM}$ .= 6,010$^{-5}$%, in the second case, $\langle ER\rangle_{JWM}$ .=5,7·10$^{-5}$%.

**Analysis of the Accuracy of the Second- and Third-Born Approximations for Calculating the Mott Correction to the Bethe–Bloch formula**

The Bethe–Bloch formula describes the average energy loss of a charged particle in matter in the first-Born approximation. The Mott correction arises in the Bethe–Bloch formula due to the difference between the Mott scattering cross section and that obtained in the first-Born approximation. The Mott correction in the second Born approximation was obtained in [17], [18] and is represented by the simple expression:

$$\Delta L_{M_2} = \frac{1}{2}\pi\alpha\beta Z. \qquad (10)$$

It is evident that the Mott correction in this approximation depends linearly on the velocity and charge of the nucleus.

The Mott correction in the third-Born approximation was obtained by Morgan and Ebi [18]. It is defined by the expression:

$$\Delta L_{M3} = \frac{1}{2}\left\{\pi\alpha\beta Z + (\alpha Z)^2\left[\pi^2/3 + 1 + \beta^2\left\{\pi^2\left(3/4 - \ln 2\right) + 0.5\left(\zeta(3) - 3\right)\right\}\right]\right\}, \qquad (11)$$

where $\zeta(3)$ is the Riemann zeta function.

As noted in [18], correction (1) is applicable only for small Z. In [20], a comparison of (1) with the difference between the Lindhard–Sørensen and Bloch corrections was performed for Z = 10, 18, and 36 as $\beta \to 1$. As shown in [21], this difference coincides with the Mott correction.

We will characterize the accuracy of the Mott correction calculation in the second or third Born approximation by the relative error:

$$\delta_{M2(3)} = \frac{\left|\Delta L_{M2(3)} - \Delta L_M\right|}{\Delta L_M}\cdot 100\%, \qquad (12)$$

where $\Delta L_M = \Delta L_{MVSTT}$ was calculated using the VSTT method developed in [32]. As the velocity increases for the second Born approximation, the relative error decreases. Table 3 lists the β values starting from which the relative error is below 5%.

**Table 3**. The value of β, starting from which $\delta_{M2} < 5\%$

| Z | 1 | 2 | 3 | 4 | 5 |
|---|---|---|---|---|---|
| β | 0,186 | 0,369 | 0,548 | 0,721 | 0,887 |

Table 4 contains the β values starting from which the error is below 10%.

**Table 4.** The value of β, starting from which $\delta_{M2} < 10\%$

| Z | 6 | 7 | 8 | 9 | 10 | 11 |
|---|---|---|---|---|---|---|
| β | 0,514 | 0,598 | 0,681 | 0,763 | 0,844 | 0,924 |

Table 5 shows the β values starting from which the error is below 20%.

**Table 5.** The value of β, starting from which $\delta_{M2} < 20\%$

| Z | 12 | 13 | 14 | 15 | 16 | 17 | 18 | 19 |
|---|---|---|---|---|---|---|---|---|
| β | 0,428 | 0,465 | 0,502 | 0,539 | 0,577 | 0,614 | 0,653 | 0,691 |
| Z | 20 | 21 | 22 | 23 | 24 | 25 | 26 | |
| β | 0,730 | 0,769 | 0,808 | 0,848 | 0,888 | 0,928 | 0,969 | |

Let us present the values of the relative error of the third-Born approximation for the specified Z and β.

**Table 6.** Relative error of the Mott correction in the third-Born approximation for certain Z and β

| | Z = 80, β = 0,95 | Z = 60, β = 0,85 | Z = 92, β = 0,99 |
|---|---|---|---|
| $\delta_{M3}$,% | 4,39 | 5,75 | 3,98 |

For Z = 1 – 3, the error is less than 3% for all β ≥ 0.1. Let's determine the β value starting from which it becomes less than 6%. The results are presented in Table 7.

**Table 7.** The β-value starting from which $\delta_{M3} < 6\%$

| Z | 4 | 5 | 6 | 7 | 8 | 9 | 10 | 11 | 12 | 13 |
|---|---|---|---|---|---|---|---|---|---|---|
| β | 0,105 | 0,121 | 0,145 | 0,168 | 0,191 | 0,214 | 0,237 | 0,259 | 0,281 | 0,302 |
| Z | 14 | 15 | 16 | 17 | 18 | 19 | 20 | 21 | 22 | 23 |
| β | 0,323 | 0,343 | 0,364 | 0,383 | 0,402 | 0,421 | 0,439 | 0,457 | 0,474 | 0,490 |
| Z | 24 | 25 | 26 | 27 | 28 | 29 | 30 | 31 | 32 | 33 |
| β | 0,506 | 0,522 | 0,537 | 0,552 | 0,566 | 0,580 | 0,594 | 0,607 | 0,619 | 0,631 |
| Z | 34 | 35 | 36 | 37 | 38 | 39 | 40 | 41 | 42 | 43 |
| β | 0,643 | 0,654 | 0,665 | 0,676 | 0,686 | 0,696 | 0,706 | 0,715 | 0,724 | 0,733 |
| Z | 44 | 45 | 46 | 47 | 48 | 49 | 50 | 51 | 52 | 53 |
| β | 0,741 | 0,750 | 0,757 | 0,765 | 0,772 | 0,780 | 0,787 | 0,793 | 0,800 | 0,806 |
| Z | 54 | 55 | 56 | 57 | 58 | 59 | 60 | 61 | 62 | 63 |
| β | 0,812 | 0,818 | 0,824 | 0,830 | 0,835 | 0,840 | 0,845 | 0,850 | 0,855 | 0,860 |
| Z | 64 | 65 | 66 | 67 | 68 | 69 | 70 | 71 | 72 | 73 |
| β | 0,865 | 0,869 | 0,874 | 0,878 | 0,882 | 0,886 | 0,890 | 0,894 | 0,898 | 0,902 |
| Z | 74 | 75 | 76 | 77 | 78 | 79 | 80 | 81 | 82 | 83 |
| β | 0,905 | 0,909 | 0,913 | 0,916 | 0,920 | 0,923 | 0,926 | 0,930 | 0,933 | 0,936 |
| Z | 84 | 85 | 86 | 87 | 88 | 89 | 90 | 91 | 92 | 93 |
| β | 0,940 | 0,943 | 0,946 | 0,949 | 0,952 | 0,956 | 0,959 | 0,962 | 0,965 | 0,969 |
| Z | 94 | 95 | 96 | 97 | 98 | 99 | 100 | | | |
| β | 0,972 | 0,975 | 0,979 | 0,982 | 0,986 | 0,989 | 0,993 | | | |

Accordingly, it is possible to determine up to what Z, inclusive, the error is below 6% for a given β.

**Table 8.** Maximum value of Z at which $\delta_{MME} < 6\%$

| β | 0,1 | 0,2 | 0,3 | 0,4 | 0,5 | 0,6 | 0,7 | 0,8 | 0,85 | 0,9 | 0,95 | 0,99 |
|---|---|---|---|---|---|---|---|---|---|---|---|---|
| $Z_{max}$ | 4 | 8 | 12 | 17 | 23 | 30 | 39 | 52 | 61 | 72 | 87 | 99 |

Although the accuracy of the third Born approximation becomes very low for large Z, the accuracy of the Mott correction remains high at high energies.

The paper [20] presents the difference between the Lindhard–Sorensen (LS) correction and the Bloch correction as β → 1. As shown in [21], this difference coincides with the Mott correction. We present our results for calculating the Mott correction using the VSTT- method [32], the difference between the LS and Bloch corrections $\Delta L_{LSMB}$ and the Mott correction in the second- and third-Born approximations, as well as the difference between the corrections from [20].

**Table 9.** Mott correction as β → 1

| Z | $\Delta L_{MBCTT}$ | $\Delta L_{LSMB}$ | $\Delta L_{M2}$ | $\Delta L_{M3}$ | $\Delta L_{LSMB}$, [20] | $\delta_{M_2}$ | $\delta_{M_2}$, [20] |
|---|---|---|---|---|---|---|---|
| 10 | 0,125241 | 0,125241 | 0,114627 | 0,125149 | 0,119 | 8,5 | 3,4 |
| 18 | 0,240932 | 0,240932 | 0,206328 | 0,240241 | 0,220 | 14,4 | 6,4 |
| 36 | 0,551343 | 0,551343 | 0,412656 | 0,551343 | 0,473 | 25,2 | 12,7 |

The last columns show the error of the second-Born approximation according to our calculations ($\delta_{M_2}$) and according to the results of [20].

Since the result of the Mott correction calculation using the VSTT method coincided with the result of calculating the difference between the Lindhard-Sorensen and Bloch corrections and was very close to the result obtained in the third Born approximation, it can be assumed that the results presented in [20] are erroneous.

The error in the second Born approximation turns out to be even higher than would be obtained from the results presented in [20].

**Second- and third-Born approximations for calculating the Mott correction for the deceleration of nuclei in antimatter or antinuclei in matter**

We will characterize the accuracy of the Mott correction calculation in the second or third Born approximation by the relative error (12).

As the velocity for the second-Born approximation increases, the relative error decreases. Table 10 lists the β-values starting from which the relative error is below 5%. The first digit in the column corresponds to negative Z, the second to positive Z.

**Table 10.** β-value starting from which $\delta_{M2} < 5\%$

| \|Z\| | 1 | 2 | 3 | 4 | 5 |
|---|---|---|---|---|---|
| β | 0,211<br>0,186 | 0,416<br>0,369 | 0,611<br>0,548 | 0,796<br>0,721 | 0,965<br>0,887 |

Table 11 contains the β values starting from which the error is below 10%.

**Table 11.** β-value starting from which $\delta_{M2} < 10\%$

| \|Z\| | 6 | 7 | 8 | 9 | 10 |
|---|---|---|---|---|---|
| β | 0,639<br>0,514 | 0,734<br>0,598 | 0,826<br>0,681 | 0,913<br>0,763 | 0,996<br>0,844 |

Table 12 contains the β values starting from which the error is below 20%.

**Table 12**. β-value starting from which $\delta_{M2} < 20\%$

| \|Z\| | 11 | 12 | 13 | 14 | 15 | 16 | 17 | 18 |
|---|---|---|---|---|---|---|---|---|
| β | 0,637<br>0,394 | 0,689<br>0,428 | 0,739<br>0,465 | 0,788<br>0,502 | 0,835<br>0,539 | 0,881<br>0,577 | 0,925<br>0,614 | 0,968<br>0,653 |

For Z ≤ -19, the error is greater than 20% for any β up to and including 0.999. The presented results show that for Z < 0, the accuracy of the second-Born approximation is lower than for Z > 0.

In the case of nuclear deceleration in matter for Z ≤ 100, for sufficiently high velocities, the relative error of the Mott correction in the third-Born approximation is less than 6% [37]. For the deceleration of antinuclei in matter (nuclei in antimatter), the error of the Mott correction at small |Z| is smaller than for the deceleration of nuclei in matter.

To compare the accuracy of the third-Born approximation for Z > 0 and Z < 0, we present a graph.

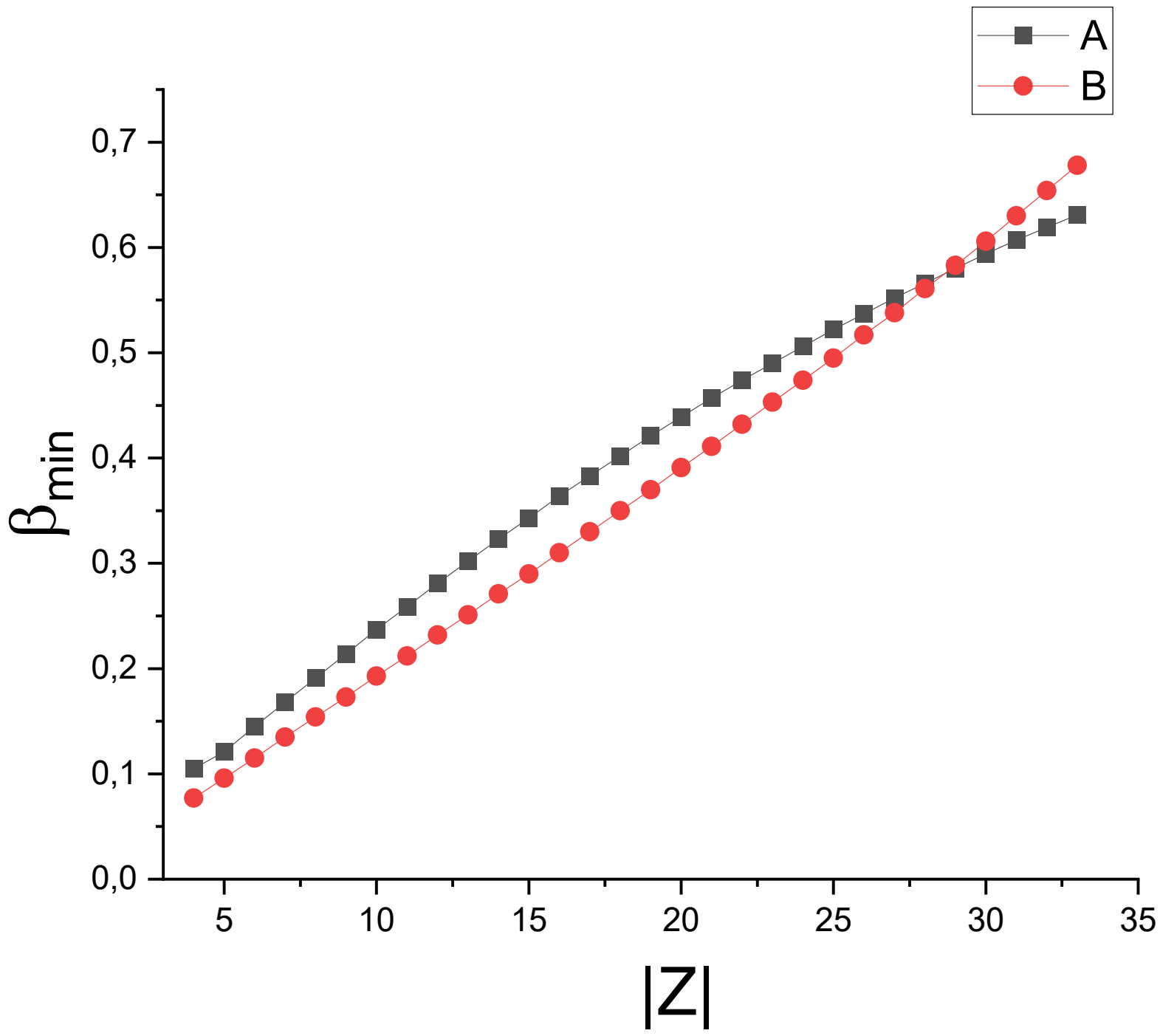

**Figure 9.** The value of β, starting from which $\delta_{M3} < 6\%$. A – Z > 0. B – Z < 0.

The graph shows that, starting with |Z|=29, the error in the third Born approximation for the Mott correction is greater for Z < 0.

For Z ≤ -52, the error in the third-Born approximation is greater than 10% for any velocity. Accordingly, for Z ≤ -65, the error in the third-Born approximation is greater than 20% for any velocity.

Note that in [18] it is indicated that at β = 0.95 there is excellent agreement between the third Born approximation for the Mott correction and the numerical value of the correction for -52 ≤ Z ≤ 80. The result of our calculations for Z = -52 and β = 0,95 gives a relative error of about 12%, which is consistent with graph 5 of the work [18].

Thus, unlike Z > 0, for Z < 0 the third Born approximation is inapplicable for Z of large absolute values. As can be seen from [18], for large Z the third-Born approximation can even give a result with the wrong sign.

**Application of Born approximations to the calculation of the atomic displacement cross section for iron, silver, and lead**

To calculate the damage to materials by high-energy electrons, the atomic displacement cross section was calculated [23]. To obtain the primary displacement cross section, a formula derived from the second-Born approximation of the differential cross section for scattering of relativistic electrons on a point nucleus is widely used [16]:

$$\sigma_{pMF}=\frac{\pi Z^2 e^4}{m^2c^4\beta^4}(1-\beta^2)\left\{\frac{1}{y}-1+\beta^2\ln(y)+\pi Z\alpha\beta\left(\frac{2}{\sqrt{y}}-2+\ln(y)\right)\right\},\ y=\frac{T_d}{T_m}. \quad (13)$$

$T_d$ is threshold energy for atomic displasement, $T_m$ is the maximum transferred energy which occurs for a head-on collision:

$$T_m=\frac{2E(E+2m_ec^2)}{Mc^2}. \quad (13\text{-}1)$$

In a number of works (see for example, [33], [34]) it was shown that with increasing Z the error of calculations according to formula (13) increases. In the paper [33], it is shown that for positrons the second-Born approximation and formula (13) are even less applicable than for electrons.

Since it was shown above that for positron scattering the first-Born approximation at large |Z| can be more accurate than the second and third one, we will consider the results of calculating the displacement cross section of atoms in the first- through third-Born approximations using specific examples.

In the first-Born approximation, the primary displacement cross section of an atom has the form (15). The expression for the primary displacement cross section of atoms in the third-Born approximation (16) was obtained in [37].

We did not perform numerical calculations of the atomic displacement cross section for the exact Mott scattering cross section. However, as shown in [40], the LQZ method for positrons provides very high accuracy. Therefore, we will use the result of calculating the atomic displacement cross section within the LQZ approximation as a reference (see [34]).

We present graphs of the primary atomic displacement cross section as a function of positron energy for copper, silver, gold, and uranium (these elements were discussed in [33]). The same threshold displacement energies as Oen's were used in the calculations.

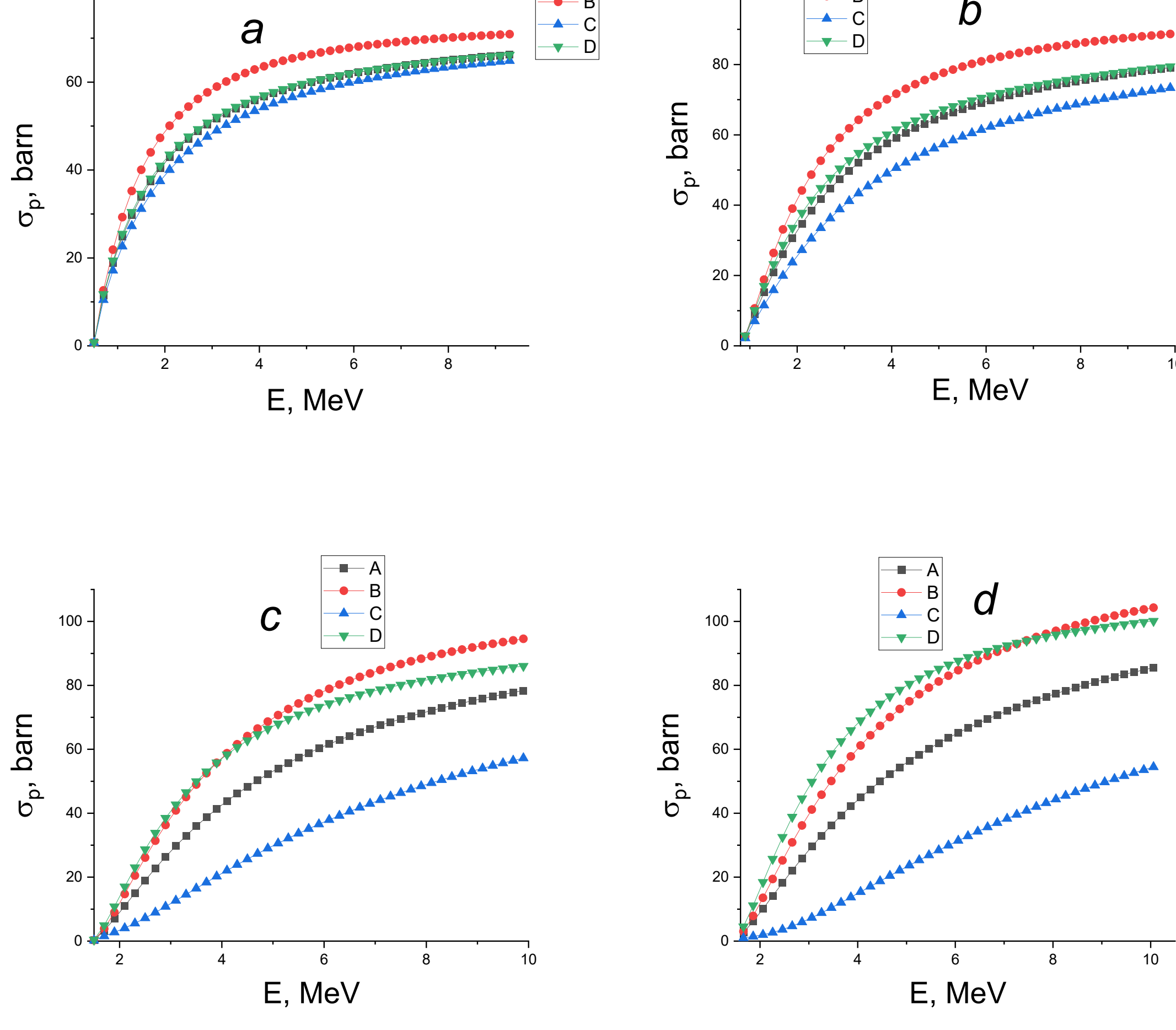

**Figure 10.** Primary atomic displacement cross section as a function of relative velocity for positron scattering. A is the LQZ method, B, C and D are first-, second- and third-Born approximations for Z = 29 (a), 47 (b), 79 (c), 92 (d).

For copper (Z = 27), the accuracy increases with increasing Born approximation number. For silver, the errors of the first and second approximations become similar. For gold, the first-Born approximation is more accurate than the second, but worse than the third. Up to approximately 4 MeV, the accuracy of the first-Born approximation is higher than that of the third. For uranium, up to approximately 7 MeV, the first approximation is more accurate than the third.

**Table 12.** Energy-averaged relative errors of the Born approximations for copper, silver, gold and uranium

| | | $\langle \delta_B \rangle$,% | $\langle \delta_{MF} \rangle$,% | $\langle \delta_{JWM} \rangle$,% |
|---|---|---|---|---|
| Cu | | 11,2 | 4,3 | 0,63 |
| Ag | | 18,4 | 13,3 | 3,9 |
| Au | | 27,7 | 40,9 | 25,2 |
| U | | 29,6 | 54,7 | 40,8 |

As Z increases, the error of all Born approximations increases, but the error of the second and third-Born approximations increases faster than that of the first. For gold, the average error of the first-Born approximation is already smaller than that of the second, and for uranium, it is smaller than that of the third. Although for the differential scattering cross section the angle- and energy-averaged error of the third-Born approximation becomes larger than the error of the second-Born approximation starting from Z = 78, for the atomic displacement cross section the third-Born approximation remains more accurate than the second-Born approximation.

**On the High-Energy Asymptotics of Various Approximations in Calculating the Atomic Displacement Cross Section**

As shown in [23, pp. 10, 11], the McKinley–Feshbach approximation at high energies for heavy elements yields results close to those obtained by numerically integrating the exact Mott cross section, whereas at lower energies, the discrepancies between the results are very significant. Similar results were obtained for platinum and gold in [34]. At first glance, this may seem surprising, since even at very high energy, the scattering cross section in the second-Born approximation differs greatly from the exact Mott cross section.

From (13), one can obtain an asymptotic expression for the atomic displacement cross section at $\frac{T_d}{T_m} \ll 1$. In this case, all terms except the first can be neglected. As a result, we get

$$\sigma_{\mathrm{pMFass}} = \frac{\pi Z^2 e^4}{m^2 c^4 \beta^4}(1-\beta^2)\frac{T_m}{T_d} = \frac{\pi Z^2 e^4}{m^2 c^4 \beta^4}(1-\beta^2)\frac{2E(E+2m_e c^2)}{T_d}\frac{1}{Mc^2}.$$

If the electron's kinetic energy is much greater than its rest energy, then the difference between β and unity in the denominator and the difference between the kinetic energy and the total energy and the term in the numerator can be neglected. Then (13) can be further simplified:

$$\sigma_{pMFass} = \frac{\pi Z^2 e^4}{m^2 c^4 \beta^4}(1-\beta^2)\frac{T_m}{T_d} = \frac{2\pi Z^2 e^4}{T_d M c^2} \qquad (14)$$

This expression in the high-energy limit for the Mott cross section was obtained in [16].

In [36] and several other works by these authors, it is claimed that formula (14) can be used for electrons with an energy of 0,9 MeV for silicon with a threshold energy of $T_d$ = 24 – 25 eV and carbon with Td = 18 eV. They justify this by the fact that the electron energy significantly exceeds the threshold energy at which atomic displacement occurs. An energy of 0.9 MeV is less than twice the rest energy of electrons. Therefore, the conditions for the applicability of asymptotics (14) are clearly not satisfied. In [36], asymptotic values of the primary atomic displacement cross section are given for carbon and silicon, □ 23 barn and □ 40 barn, respectively. Calculation using formula (13) leads to results of 23.3 barn for carbon and 40.7 barn and 39,0 barn for silicon for $T_d$ = 24 and 25 eV, respectively. Calculation using formula (13) yields results of 22,1 barns, 28,2 barns, and 26.5 barns, respectively. The difference between the asymptotic result and the result using formula (13) for carbon is approximately 5%. For silicon, it is 44–47%. Moreover, the electron energy in the case of silicon exceeds the threshold energy causing atomic displacement by 3,5–3,6 times. Thus, this is clearly insufficient for the applicability of asymptotes (14).

We calculated the cross section for the primary displacement of an atom for a number of elements using the second- and third-Born approximations. The results are presented in the Table. Also presented the quantities $\frac{T_m}{T_d}, \frac{E}{E_d}$ that are the ratio of the electron kinetic energy to the threshold kinetic energy of the electron at which displacement occurs, the result of the calculation using formula (14), and the relative error of this formula with respect to formula (13) δ, respectively. Errors exceeding 1% are shown in bold. For comparison, the table also includes the result of numerical calculations of the cross section from [23].

**Table 13.** Cross section of primary displacement of an atom

| Z = 1, M = 1,008, $T_d$ = 4 эВ, E = 18,3 МэВ | | | | | | |
|---|---|---|---|---|---|---|
| $\sigma_{pMF}$ | $\sigma_{pJWM}$ | $\sigma_{pMFass}$ | δ, % | $T_m/T_d$ | $E/E_d$ | $\sigma_{pO}$ [1] |
| 34,7151 | 34,7151 | 34,6883 | 0,077 | $1,88\cdot10^{5}$ | $9,98\cdot10^{3}$ | 34,74 |
| Z = 1, M = 1,008, $T_d$ = 96 эВ, E = 169 МэВ | | | | | | |
| 1,44541 | 1,44541 | 1,44535 | 0,004 | $6,38\cdot10^{5}$ | $3,99\cdot10^{3}$ | 1,45 |
| Z = 2, M = 4,003, $T_d$ = 4 эВ, E = 72,4 МэВ | | | | | | |
| 34,9444 | 34,9444 | 34,9396 | 0,014 | $7,13\cdot10^{5}$ | $9,99\cdot10^{3}$ | 34,98 |
| Z = 2, M = 4,003, $T_d$ = 96 эВ, E = 304 МэВ | | | | | | |
| 1,45596 | 1,45596 | 1,45582 | 0,010 | $5,18\cdot10^{5}$ | 1995 | 1,46 |
| Z = 4, M = 9,013, $T_d$ = 20 эВ, E = 305 МэВ | | | | | | |
| 12,4164 | 12,4164 | 12,4144 | 0,016 | $1,11\cdot10^{6}$ | $3,99\cdot10^{3}$ | 12,44 |
| Z = 4, M = 9,013, $T_d$ = 24 эВ, E = 362 МэВ | | | | | | |
| 10,3468 | 10,3468 | 10,3453 | 0,015 | $1,30\cdot10^{6}$ | 3998 | 10,36 |
| Z = 6, M = 12,011, $T_d$ = 32 эВ, E = 304 МэВ | | | | | | |
| 13,1048 | 13,1048 | 13,1002 | 0,035 | $5,18\cdot10^{5}$ | 1994 | 13,13 |
| Z = 6, M = 12,011, $T_d$ = 36 эВ, E = 338 МэВ | | | | | | |
| 11,6485 | 11,6486 | 11,6446 | 0,034 | $5,69\cdot10^{5}$ | 1999 | 11,67 |
| Z = 13, M = 26,98, $T_d$ = 12 эВ, E = 523 МэВ | | | | | | |
| 73,0389 | 73,0392 | 73,0074 | 0,043 | $1,82\cdot10^{6}$ | 3998 | 73,29 |
| Z = 13, M = 26,98, $T_d$ = 16 эВ, E = 168 МэВ | | | | | | |
| 54,8363 | 54,8379 | 54,7556 | 0,15 | $1,41\cdot10^{5}$ | 995 | 55,01 |
| | Z = 29, M = 63,54, $T_d$ = 16 эВ, E = 173 МэВ | | | | | |
| 116,273 | 116,304 | 115,700 | 0,49 | $6,36\cdot10^{5}$ | 500 | 116,66 |
| Z = 29, M = 63,54, $T_d$ = 20 эВ, E = 206 МэВ | | | | | | |
| 92,9919 | 93,0142 | 92,5598 | 0,46 | 72E3 | 499 | 93,3 |
| Z = 50, M = 118,7, $T_d$ = 12 эВ, E = 450 МэВ | | | | | | |
| 246,471 | 246,527 | 245,477 | 0,40 | $3,06\cdot10^{5}$ | 999 | 243,87 |
| | Z = 50, M = 118,7, $T_d$ = 96 эВ, E = 184 МэВ | | | | | |
| 31,4575 | 31,5911 | 30,6847 | **2,46** | $6,14\cdot10^{3}$ | 99,5 | 31,33 |
| Z = 82, M = 207,21, $T_d$ = 12 эВ, E = 340 МэВ | | | | | | |
| 382,567 | 383,13 | 378,215 | **1,14** | $1,00\cdot10^{5}$ | 500 | 366,88 |
| Z = 82, M = 207,21, $T_d$ = 96 эВ, E = 257 МэВ | | | | | | |
| 49,1778 | 49,6897 | 47,2769 | **3,87** | 7157 | 99,8 | 47,96 |

| Z = 92, M = 238, $T_d$ = 32 эВ, E = 288 МэВ | | | | | | |
|---|---|---|---|---|---|---|
| 159,475 | 160,359 | 155,436 | **2,53** | $23{,}5\cdot10^3$ | 200 | 153,07 |
| Z = 92, M = 238, $T_d$ = 36 эВ, E = 310 МэВ | | | | | | |
| 141,706 | 142,474 | 138,165 | **2,50** | $24{,}2\cdot10^3$ | 200 | 135,97 |
| Z = 94, M = 242, $T_d$ = 4 эВ, E = 166 МэВ | | | | | | |
| 1298,05 | 1301,7 | 1276,68 | **1,65** | $61{,}5\cdot10^3$ | 499 | 1228,85 |
| Z = 94, M = 242, $T_d$ = 96 эВ, E = 281 МэВ | | | | | | |
| 55,6319 | 56,3762 | 53,1952 | **4,38** | $7{,}3\cdot10^3$ | 99,7 | 53,66 |
| Z = 99, M = 254, $T_d$ = 4 эВ, E = 172 МэВ | | | | | | |
| 1372,75 | 1376,95 | 1349,21 | **1,71** | $62{,}9\cdot10^3$ | 497 | 1290,06 |
| Z = 99, M = 254, $T_d$ = 96 эВ, E = 289 МэВ | | | | | | |
| 58,9237 | 59,7914 | 56,2172 | **4,59** | $7{,}38\cdot10^3$ | 99,7 | 56,56 |

Even at $E/E_d = 100$, the error when using simplified formula (18) for heavy elements exceeds 4%. At $E/E_d = 500$, the error for heavy elements exceeds 1%.

The specific energies at which the asymptotic expression for the atomic displacement cross section can be used must be assessed specifically for each specific case.

For a number of examples for which specific $T_d$ values are used in various sources, we calculated the electron energy E at which the difference between the cross section obtained using asymptotic formula (14) and that obtained using formula (13) would be less than 1%. The results are presented in the table 14. The ratio of the electron kinetic energy to the threshold kinetic energy $\frac{E}{E_d}$ at which displacement occurs is also given. The threshold kinetic energy can be found from (2) by equating the maximum energy (13') to $T_d$:

$$E_d = \sqrt{m^2c^4 + \frac{T_d M c^2}{2}} - mc^2. \tag{15}$$

**Table 14.** Cross section of primary displacement of an atom

| **Z = 4, M = 9,012** | | | $T_d$, эВ | E, МэВ | $E/E_d$ | 29[9] | 71,7 | 168,3 |
|---|---|---|---|---|---|---|---|---|
| $T_d$, эВ | E, МэВ | $E/E_d$ | 14 [6] | 11,4 | 76,0 | **Z = 24, M = 63,54** | | |
| 22[6] | 2,78 | 33,3 | 16 [9] | 3,393 | 20,1 | $T_d$, эВ | E, МэВ | $E/E_d$ |
| **Z = 5, M = 10,81** | | | **Z = 14, M = 28,09** | | | 22[9] | 66,76 | 175,7 |
| $T_d$, эВ | E, МэВ | $T_d$, эВ | $T_d$, эВ | E, МэВ | $E/E_d$ | 24[6] | 69,71 | 171,4 |
| 28 [6] | 4,89 | 39,7 | 13 | 15,4 | 105,7 | **Z = 25, M = 54,94** | | |
| **Z = 6, M = 12,01** | | | 24 [7] | 18,7 | 75,6 | $T_d$, эВ | E, МэВ | $E/E_d$ |
| $T_d$, эВ | E, МэВ | $E/E_d$ | 25[7] | 18,92 | 73,9 | 19[6] | 67,25 | 190,3 |
| 18 [7] | 2,92 | 32,3 | 33[8] | 20,6 | 64,1 | **Z = 26, M = 55,85** | | |
| 33 [6] | 5,61 | 35,8 | 100 [10] | 11,17 | 15,1 | $T_d$, эВ | E, МэВ | $E/E_d$ |
| **Z = 8, M = 16,00** | | | **Z = 15, M = 30,97** | | | 16 [9] | 65,43 | 209,7 |
| $T_d$, эВ | E, МэВ | $E/E_d$ | $T_d$, эВ | E, МэВ | $E/E_d$ | 20[6] | 73,11 | 196,0 |
| 16,5 [8] | 3,064 | 28,2 | 12 | 18,36 | 124,1 | **Z = 27, M = 58,93** | | |
| 100[8] | 11,92 | 24,2 | Z = 22, M = 47,87 [9] | | | $T_d$, эВ | E, МэВ | $E/E_d$ |
| **Z = 12, M = 24,31** | | | $T_d$, эВ | E, МэВ | $E/E_d$ | 22[6] | 82,54 | 197,0 |
| $T_d$, эВ | E, МэВ | $E/E_d$ | 19 | 53,05 | 167,6 | 23[9] | 84,4 | 194,6 |

| 14 | 2,872 | 21,0 | **Z = 23, M = 50,94** | | | **Z = 30, M = 65,38** | | |
|---|---|---|---|---|---|---|---|---|
| 25 | 4,834 | 21,3 | $T_d$, эВ | E, МэВ | *E/E<sub>d</sub>* | $T_d$, эВ | E, МэВ | $E/E_d$ |
| **Z = 13, M = 26,98** | | | 26 [6] | 67,9 | 159,4 | 52 [8] | 152 | 179,4 |

$E/E_d$ varies for the cases considered from 15,1 to 209,7. The case of aluminum is interesting: when going from $T_d$ = 14 eV to $T_d$ = 16 eV, E/Ed drops from 76 to 20,1. This is due to the fact that at $T_d$ = 14 eV, the error drops below 1% already at an energy of 2.98 MeV, but at 7,9 MeV, it becomes higher than 1% in absolute value and reaches a maximum absolute value of approximately 1.024%. For $T_d$ = 16 eV, after 3,393 MeV, the error modulus remains less than 1%.

For the elements considered in the Table 14, the second Born approximation has good accuracy, and it can be assumed that the error of the asymptotic expression of less than 1% for the second-Born approximation ensures a correspondingly small error value for the exact displacement cross section. It should be noted that at energies greater than 10 MeV, the size of the nucleus begins to play a role, and the use of the Mott scattering cross section obtained for a point nucleus is not accurate.

**Accuracy of Born Approximations for Calculating Energy-Loss Straggling**

Energy-loss straggling (ELS) is the dispersion of particle energy losses as they pass through matter, see formula (15):

$$W = \langle(\delta E - \langle\delta E\rangle)^2\rangle, \tag{15}$$

where δE is the energy loss.

The ELS for charged particles plays an important role in both fundamental research and many practical applications in physics and other disciplines.

The main contributions to the ELS come from small momentum transfers, large momentum transfers, and the exchange of particle charge in matter. The resulting ELS-value can be found as the sum of the individual contributions [38]:

$$W = W_{high-Q} + W_{low-Q} + W_{ch.-ex.} \tag{16}$$

According to [39], the first term can be represented as

$$\frac{dW_{high-Q}}{dx} = 4\pi Z^2 e^4 n Z_2 \gamma^2 X, \gamma = \frac{1}{\sqrt{1-\beta^2}}, \beta = \frac{v}{c}. \tag{17}$$

Here X is a dimensionless quantity. In the first-Born approximation for a relativistic point charge, this quantity, according to [39], has the form

$$X_B = 1 - \beta^2/2. \tag{18}$$

In [39] and [40], methods for calculating the X-value for the deceleration of a point nucleus were independently developed. As was shown [40], for the nuclei deceleration of in matter at large Z, the first-Born approximation yields a large error. The exact X-value can exceed that obtained in (18) by more than 5 times. However, for Z < 0 (the deceleration of antinuclei in matter or nuclei in antimatter), the error in the first-Born approximation is much smaller. For example, at Z = -118, the difference is less than 25%.

To obtain expressions for X in second- and third-Born approximations, we use the formula [40]:

$$X = 0,5\int_0^{\pi} R(\theta)\sin\theta d\theta. \tag{19}$$

From here, the following expressions in second- and third-Born approximations, can be obtained, respectively:

$$X_{MF} = 1 - \beta^2/2 + \frac{\pi Z\alpha\beta}{6}. \tag{20}$$

$$X_{JWM} = 1 - \beta^2/2 + \frac{\pi Z\alpha\beta}{6} + \frac{(Z\alpha)^2}{144}(114 + 8\pi^2 + 3\beta^2\{2\pi^2[17 - 24\,ln\,2] + 3[8\zeta(3) - 19]\}).$$

The expression of relative errors for X-value can be represented as

$$\delta X = \frac{X - X_V}{X_V}. \tag{21}$$

The value of X was calculated using the Voskresenskaya method (V-method) [41, 42]. Then, the arithmetic mean modules of relative errors (average relative errors) were calculated for 26 different speeds (from 0.1 *c* to 0.999 *c*), according to the expression

$$\langle \delta X \rangle = \frac{1}{26}\sum_{i=1}^{26}\left|\delta X_i\right|. \tag{22}$$

The results are shown in the following Figure (11).

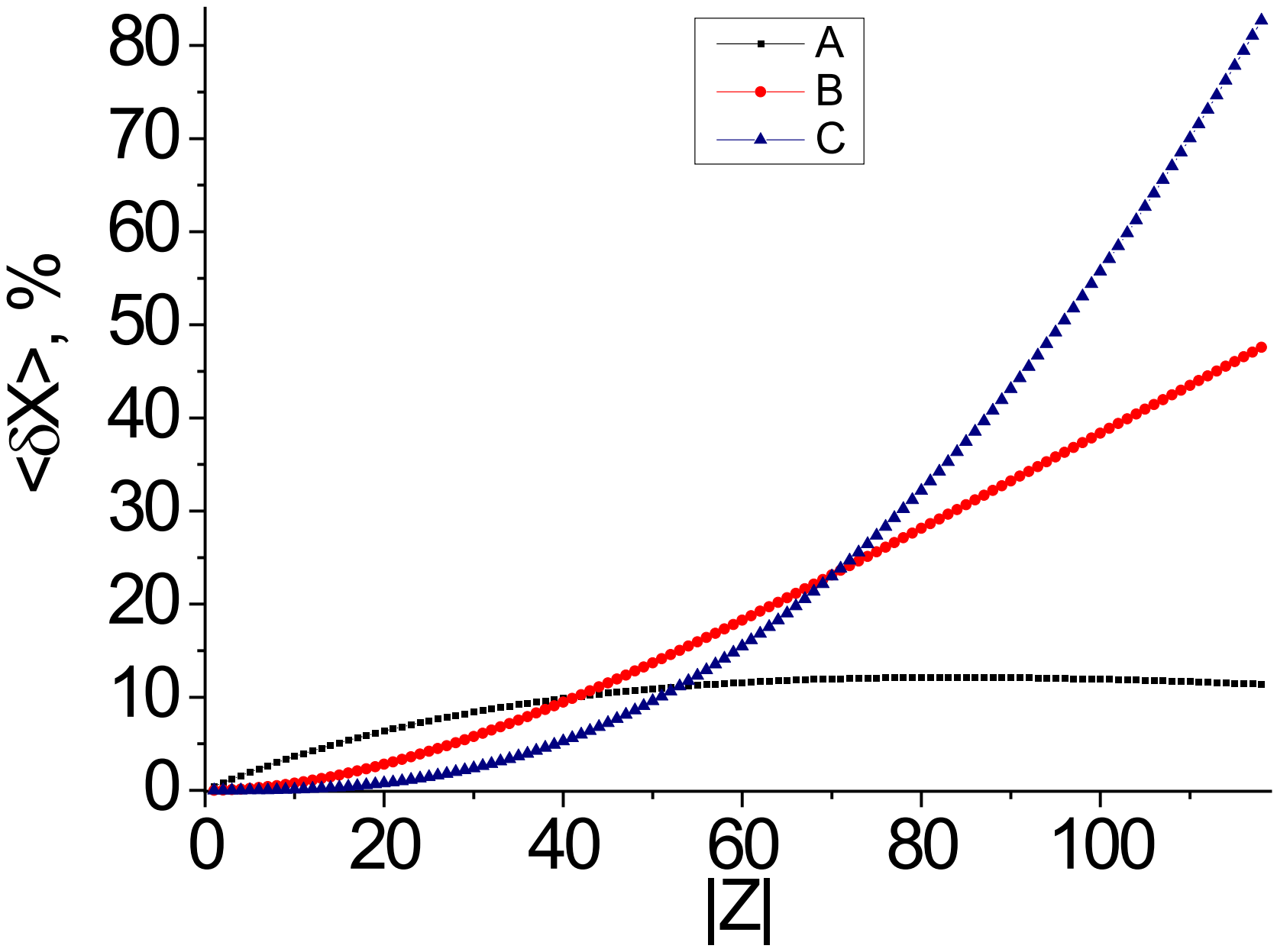

**Figure 11.** Dependence on |Z| during the deceleration of antinuclei in matter (or nuclei in antimatter): A, B, and C are $\langle \delta X_B \rangle$, $\langle \delta X_{MF} \rangle$, and $\langle \delta X_{JWM} \rangle$, respectively.

For small |Z|, the errors in the second- and third-Born approximations are very small and significantly smaller than the error in the first-Born approximation. Starting at |Z| = 42, the second- Born approximation becomes, on average, less accurate than the first. For the third Born approximation, this occurs starting at |Z| = 53.

The increase in the error in the first-Born approximation with increasing |Z| is non-monotonic. The velocity-averaged error reaches a maximum at |Z| = 84 and then slowly decreases.

The graphs show the values of X as a function of γ - 1 for Z = -15, Z = -41, and Z = -118. A is the exact calculation using V-method [41,42]; B, C, and D are calculations within the first, second, and third-Born approximations, respectively.

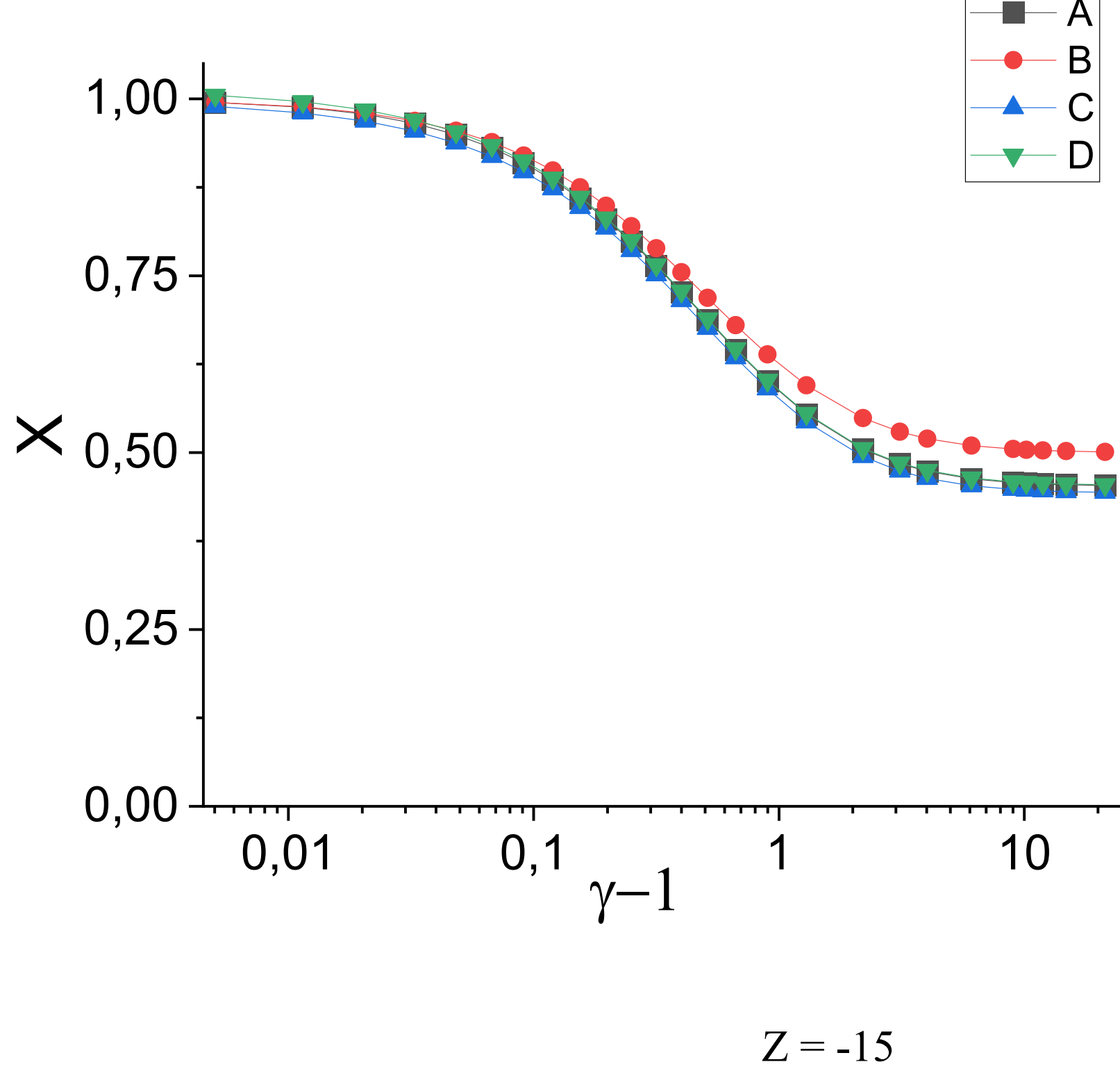

**Figure 12.** X values as a function of γ - 1 for Z = -15

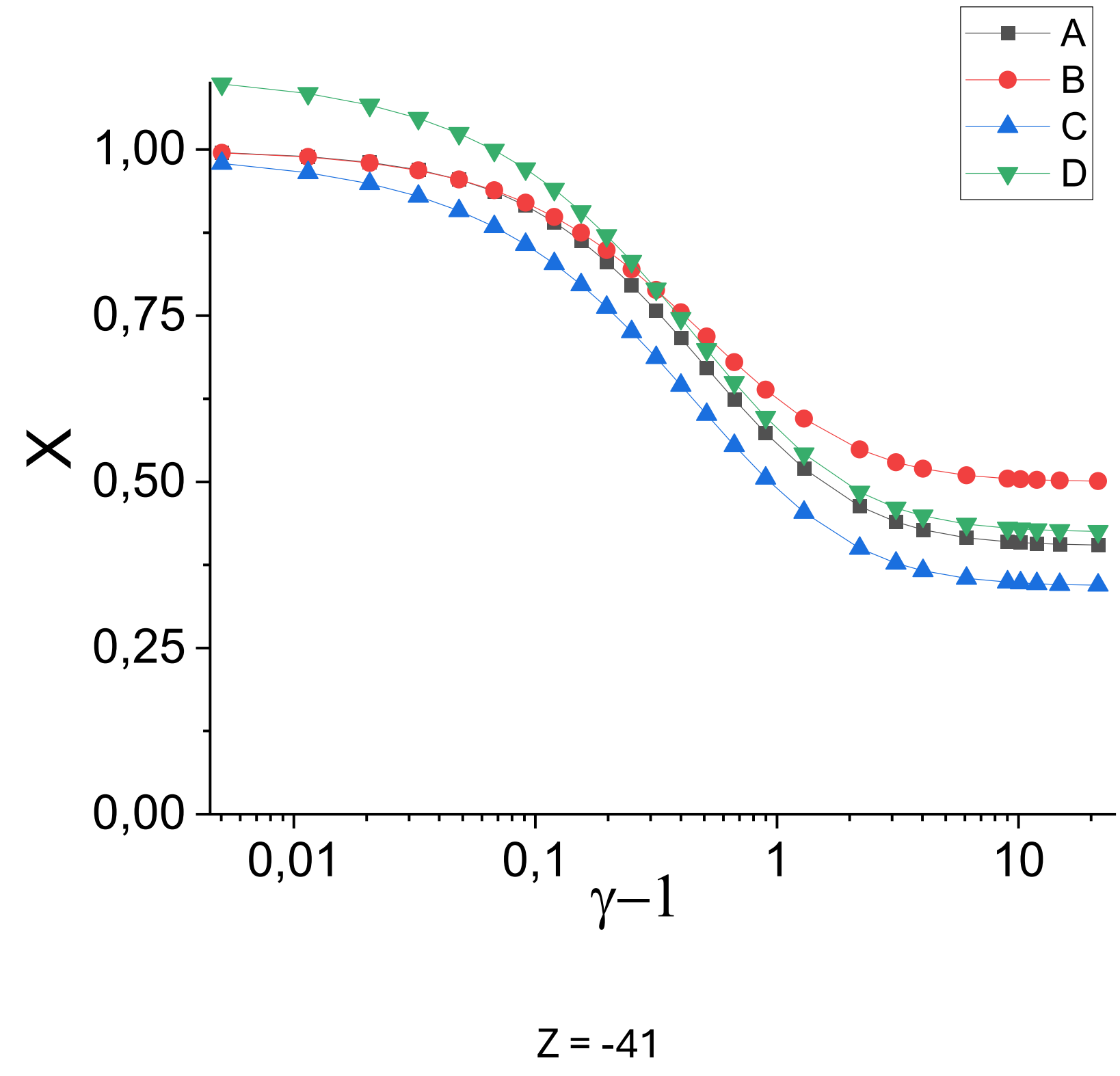

**Figure 13.** X values as a function of γ - 1 for Z = -41

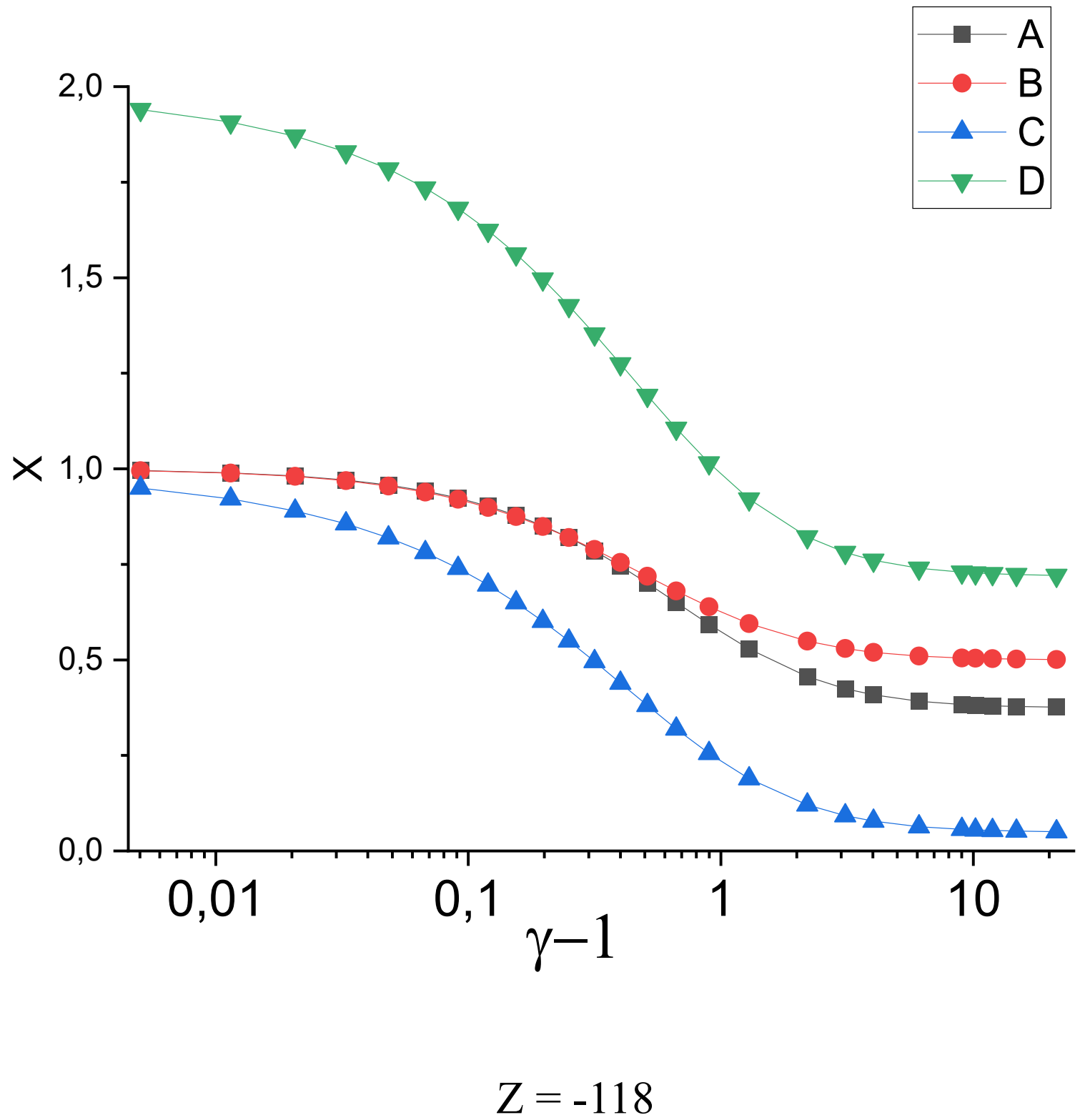

Z = -118

**Figure 14.** Graph of X versus $\gamma - 1$ for Z = -118

.

The first-Born approximation for all Z < 0 provides good accuracy at low velocities. The third Born approximation, like the Mott correction, yields incorrect asymptotics for heavy elements at low velocities, and the error decreases with increasing velocity. However, unlike the Mott correction, the error decreases non-monotonically. Here is a plot of X versus $\gamma - 1$ for Z = -53.

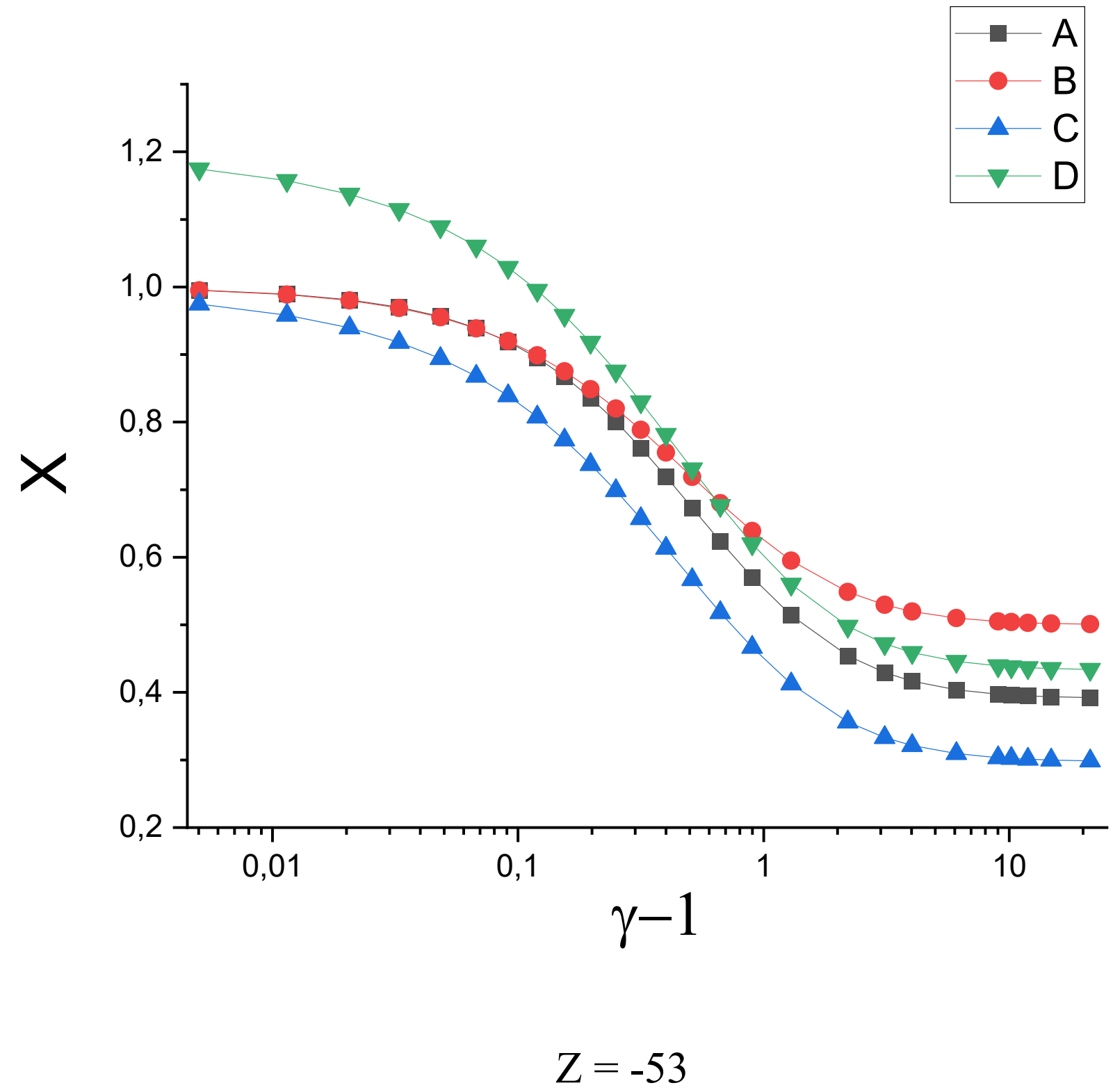

Z = -53

**Figure 15.** Graph of X versus γ – 1 for Z = -53.

For the same average error, the first-Born approximation yields a smaller error up to approximately β = 0,75. For the entire relative velocity range considered, from 0,1 *c* to 0,999 *c*, the third-Born approximation is more accurate than the first up to and including |Z| = 5.

**Table 15.** Value of |Z| (up |Z| = 8 to |Z| = 65) including which the error in the third-Born approximation for calculating energy straggling is smaller than the error in the first-Born approximation at the velocities β up 0,15 to 0,9

| β | 0,15 | 0,2 | 0,25 | 0,3 | 0,4 | 0,5 | 0,6 | 0,7 | 0,8 | 0,9 |
|---|---|---|---|---|---|---|---|---|---|---|
| \|Z\| | 8 | 10 | 13 | 16 | 22 | 29 | 36 | 44 | 54 | 65 |

Table 15 lists the |Z|-values (up |Z| = 8 to |Z| = 65) including which the error in the third-Born approximation for calculating energy straggling is smaller than the error in the first-Born approximation (at relative velocities β greater than or equal to the given in this Table).

## Conclusion

This paper primarily considers the Born approximations for electron scattering by nuclei with charge numbers ranging from 1 to 100, and the following results were obtained.

- The relative error in calculating the NMS reaches 5% in absolute value for the first-Born approximation at Z = 5; for the second one at Z = 18, and for the third at Z = 31 at an electron energy of 10 MeV. Up to an energy of 25 keV, the accuracy of the nonrelativistic approximation (Rutherford cross section) is higher than that of the first- and second-Born approximations.
- The error of the first-Born approximation increases with increasing velocity, the error of the second-Born approximation depends weakly on velocity (a maximum is visible in the dependence graph for large Z-values), and the error of the third-Born approximation decreases with increasing velocity. The relative error averaged over angles and velocities reaches 5% for Z = 11, Z = 26, and Z = 42 for the first, second, and third-Born approximations, respectively.
- It is shown that when using the coefficients for LQZ from [30] for $Z \leq 3$, the error in the third-Born approximation is lower than the error of the LQZ method.
- It is also shown that the error in the Born approximations for calculating the Mott correction increases with increasing Z and the relative error decreases with increasing velocity. In addition, we have shown that the relative error when using the second Born approximation for calculating the Mott correction is higher than when calculating the differential scattering cross section.
- We also found that the accuracy of the Mott correction in the third-Born approximation remains high at high energies.
- One of the main results of the work is the obtaining through special functions of an analytical expression for the cross section of the primary displacement of an atom in the third-Born approximation. It is shown that the error in calculating the primary atomic displacement cross section for iron is less than 5% at energies above 1,93 MeV and it is above 0,380 MeV for the second- and third-Born approximations, respectively. For silver, the error is less than 5% at energies above 3,72 MeV and 1,41 MeV. We also showed that the error in the third-Born approximation is in some cases, higher than that in the second-Born approximation. The reasons for this are analyzed.
- It is shown that even when the threshold energy is exceeded by a factor of 100, the error in the simplified formula for the second-Born approximation can exceed 4% for heavy elements.
- For several examples, it is showed that, starting from the electron energy, the difference between the cross section obtained using the asymptotic formula and that obtained using the McKinley–Feshbach formula will be less than 1%. It has been established that the ratio of the electron energy to the threshold kinetic energy of the electron, at which displacement occurs, ranges from 15,1 to 209,7 for the cases considered. This demonstrates that great caution is required when applying the simplified formula for calculating the primary atomic displacement cross section.
- It is shown that for $Z < 0$ (positron scattering, deceleration of nuclei in antimatter or anti-nuclei in matter), the first Born approximation can be more accurate than the second and third approximations under certain conditions.

In this work we also obtained the following main results for the energy loss straggling.

- We calculated the relative error $\delta X$ (21) of the value X from the expression for straggling (17) in the second-, third-Born and some other approximations.
- Besides, we obtained the arithmetic mean modules of the $\delta X$ for 118 different |Z|-values.
- We also obtained the X-values as a function of γ-1 for Z = -15, -41, -53, and -118 within the first-, second-, and third-Born approximations.
- Accuracy of Born approximations in calculating the energy loss straggling for the entire β range, from 0,1 s to 0,999 s, and |Z|-range from 8 to 118 was analyzed.